\documentclass[12pt]{article}

\usepackage{epsf,amsfonts,amssymb,epsfig,amsmath,amscd}

\renewcommand{\text}[1]{#1}
\newcommand{\bref}[1]{(\ref{#1})}

\newcommand{\be}{\begin{equation}}
\newcommand{\ee}{\end{equation}}
\newcommand{\ben}{\begin{displaymath}}
\newcommand{\een}{\end{displaymath}}
\newcommand{\bea}{\begin{eqnarray}}
\newcommand{\eea}{\end{eqnarray}}
\newcommand{\bean}{\begin{eqnarray*}}
\newcommand{\eean}{\end{eqnarray*}}
\newcommand{\nn}{\nonumber \\}
\newcommand{\ba}{\begin{array}}
\newcommand{\ea}{\end{array}}
\newcommand{\bi}{\begin{itemize}}
\newcommand{\ei}{\end{itemize}}

\def\a{\alpha}

\def\b{\beta}
\def\g{\gamma}

\def\G{\Gamma}

\def\e{\epsilon}

\def\w{\omega}

\def\CR{\mathcal{R}}
\def\CM{\mathcal{M}}

\newcommand{\sac}{\, , \qquad}

\def\undos{{1\over 2}}

\DeclareMathOperator{\dP}{dP}

\newcommand{\calp}{\mbox{${\cal P}$}}

\newcommand{\pa}{\partial}

\newcommand{\bbR}{{\mathbb{R}}}
\newcommand{\bbZ}{{\mathbb{Z}}}

\newcommand{\bbC}{{\mathbb{C}}}
\newcommand{\bbP}{{\mathbb{P}}}

\DeclareMathOperator{\re}{Re}
\DeclareMathOperator{\im}{Im}

\def\tP{\tilde{P}}
\def\te{\tilde{e}}
\def\1f{f_1^{1/2}}
\def\2f{f_2^{1/2}}
\def\4f{f_4^{1/2}}

\newcommand{\Vol}[1]{\mbox{vol}_{#1}}

\begin{document}


\makeatletter
\renewcommand{\theequation}{\thesection.\arabic{equation}}
\@addtoreset{equation}{section}
\makeatother

\baselineskip 18pt

\begin{titlepage}

\vfill

\begin{flushright}
Imperial/TP/2006/JG/02\\
\end{flushright}

\vfill

\begin{center}
   \baselineskip=16pt
   {\Large\bf  New Supersymmetric $AdS_3$ Solutions}
   \vskip 2cm
      Jerome P. Gauntlett, Ois\'{\i}n A. P. Mac Conamhna, Toni Mateos \\
      and Daniel Waldram
   \vskip .6cm
      \begin{small}
      \textit{Theoretical Physics Group, Blackett Laboratory, \\
        Imperial College, London SW7 2AZ, U.K.}
        \end{small}\\*[.6cm]
      \begin{small}
      \textit{The Institute for Mathematical Sciences, \\
        Imperial College, London SW7 2PE, U.K.}
        \end{small}
\vskip 2cm
      \begin{small}
      \textit{Dedicated to the memory of Andrew Chamblin}
        \end{small}
   \end{center}

\vfill

\begin{center}
\textbf{Abstract}
\end{center}
We construct infinite new classes of supersymmetric solutions of
$D=11$ supergravity that are warped products of $AdS_3$ with an
eight-dimensional manifold $\CM_8$ and have non-vanishing four-form
flux. In order to be compact, $\CM_8$ is constructed as an $S^2$
bundle over a six-dimensional manifold $B_6$ which is either
K\"ahler-Einstein or a product of K\"ahler-Einstein spaces. In the
special cases that $B_6$ contains a two-torus, we also obtain new
$AdS_3$ solutions of type IIB supergravity, with constant dilaton and
only five-form flux. Via the AdS-CFT correspondence the solutions with
compact $\CM_8$ will be dual to two-dimensional conformal field
theories with $N=(0,2)$ supersymmetry. Our construction can also
describe non-compact geometries and we discuss examples in
type IIB which are dual to four-dimensional $N=1$ superconformal
theories coupled to string-like defects.

\begin{quote}

\end{quote}

\vfill

\end{titlepage}
\setcounter{equation}{0}


\tableofcontents


\section{Introduction}


In this paper we will construct infinite new classes of supersymmetric
solutions of $D=11$ supergravity and type IIB supergravity that
contain $AdS_3$ factors and compact internal spaces. Via the AdS-CFT
correspondence~\cite{Maldacena:1997re} these solutions are dual to
two-dimensional conformal field theories with $N=(0,2)$
supersymmetry. Note that a subclass of the type IIB solutions was
already discussed in~\cite{GMMW2}.

Our construction is directly inspired by the supersymmetric solutions
found in~\cite{GMSW}.  Recall that these solutions
are warped products of $AdS_5$ with a six-dimensional manifold $M_6$ and
are all dual to $N=1$ supersymmetric conformal field theories in
four dimensions. These solutions were found in two steps. The first
step consisted of classifying the most general supersymmetric
solutions of this form using G-structure techniques.
The second step involved imposing a suitable ansatz on $M_6$, namely
that $M_6$ is a complex manifold, and then showing that all such
compact $M_6$ could be constructed in explicit form. Specifically
$M_6$ is an $S^2$ bundle over a four-dimensional base which is either
(i) K\"ahler-Einstein with positive curvature i.e.
$S^2\times S^2$, $CP^2$ or a del-Pezzo $\dP_k$ with $k=3,\dots, 8$,
or (ii) a product: $S^2 \times S^2$, $S^2\times H^2$ or $S^2\times
T^2$, each factor with
its constant curvature metric.
The last example with base $S^2\times T^2$ is related, via dimensional
reduction and T-duality, to a family of type IIB solutions $AdS_5\times Y^{p,q}$
where $Y^{p,q}$ are new Sasaki-Einstein metrics on $S^2\times S^3$~\cite{GMSW2}.  These
five-dimensional Sasaki-Einstein metrics, and their generalisations
\cite{Cvetic:2005ft}, have been receiving much attention because the
dual conformal field theories can be identified
\cite{gauge}--\cite{gauge5}. It is an important outstanding issue to
elucidate the conformal field theories dual to the other M-theory
solutions found in~\cite{GMSW}.

It is natural to try and repeat the successful constructions of~\cite{GMSW}
in different contexts. In~\cite{Gauntlett:2005ww} a complete
classification of the most general supersymmetric solutions of type
IIB supergravity consisting of warped products of $AdS_5$ with a
five-dimensional manifold $X_5$ was successfully carried
out\footnote{For some recent further developments see
 ~\cite{ruben}.}. The Pilch-Warner solution~\cite{Pilch:2000ej} was
recovered using this formalism, but, as yet, it is unclear what
additional ansatz one should impose upon $X_5$ in order to be able to
construct new explicit solutions. Furthermore, a detailed
classification of $AdS_5$, $AdS_4$, and $AdS_3$ solutions of $D=11$
supergravity with various amounts of supersymmetry and vanishing
electric four-form flux was given in~\cite{GMMW1}. While various known
solutions were recovered, it again proved difficult to find new
classes of solutions.

In this paper we will construct solutions of $D=11$ supergravity that are the
warped product of $AdS_3$ with an eight-dimensional manifold $\CM_8$ which
are dual to conformal field theories with $N=(0,2)$ supersymmetry.
The solutions, which have non-vanishing electric four-form flux,
do not fall within the classes studied in~\cite{GMMW1}, so require an
appropriate generalisation, which is discussed in detail in
appendix~\ref{sec:appendix}. By considering a suitable ansatz, we will
then show that we can indeed find infinite new classes of  explicit
$AdS_3$ solutions of $D=11$ supergravity. We
find solutions in which $\CM_8$ is an $S^2$ bundle over a
six-dimensional base space $B_6$ which is either a six dimensional
K\"ahler-Einstein space, $KE_6$, a product of a four- and a
two-dimensional K\"ahler-Einstein spaces, $KE_4\times KE_2$, or a
product of a three two-dimensional K\"ahler-Einstein spaces,
$KE_2\times KE_2\times KE_2$. There are various possibilities for the
signs of the curvature of the K\"ahler-Einstein spaces, as we shall
see. If these backgrounds are to be bona fide M-theory solutions one
must also ensure that the four-form flux is quantised. While we do not
consider this in general, the expectation is that this will restrict
most if not all of the parameters in the solutions to discrete
values. We show how that this is indeed the case in a particular class
of solutions.

In the special case when $\CM_8$ is an $S^2$ bundle over a $KE_4\times T^2$
or a $KE_2\times KE_2\times T^2$ base space, we can dimensionally
reduce and then T-dualise to give solutions of type IIB supergravity
(this is analogous to how the $AdS_5\times Y^{p,q}$ solutions were
found in~\cite{GMSW}).  These type IIB solutions are warped products
of $AdS_3$ with a seven manifold $\CM_7$, have constant dilaton and
non-vanishing five-form flux.  For those arising from $KE_4\times
T^2$, we will show that there are two families of
regular solutions for any positively curved $KE^+_4$. One family was
first presented and analysed in some detail in the type IIB context
in~\cite{GMMW2}. Of the second family, taking the special case where
$KE^+_4=CP^2$, the IIB solution describes the near horizon limit of
D3-branes wrapping a holomorphic Riemann surface embedded in a
Calabi-Yau four-fold that was first constructed in~\cite{naka}
(generalising a construction of Maldacena and
N\'u\~nez~\cite{mn}). For this family we calculate the central charge
of the dual CFT. Remarkably this is integral independent of the choice
of $KE^+_4$. The new $D=11$ solutions arising from a $KE_2\times
KE_2\times T^2$ base space give rise to new type IIB solutions
generalising these solutions and also those presented in~\cite{GMMW2}.

Most of the paper will focus on solutions in which the internal space
is compact. However, generically our ansatz also includes non-compact
solutions. We will briefly discuss a class of these in the type IIB
context, that is of the form of a warped product $AdS_3\times\CM_7$
with non-compact $\CM_7$. These solutions include backgrounds that
can be interpreted as the back-reacted geometry of probe D3-branes
in $AdS_5\times S^5$ with world-volume $AdS_3\times S^1$, and preserving
$\tfrac{1}{16}$th of the supersymmetry. The corresponding dual field theory
is $N=4$ super Yang-Mills coupled to string-like defects preserving
the superconformal group in two dimensions. More
generally, the $S^5$ factor can be replaced by a Sasaki-Einstein
manifold, and the back-reacted geometry corresponds to some four-dimensional
$N=1$ SCFT coupled to string-like defects.

The plan of the rest of the paper is as follows. We start by
motivating the ansatz for the eleven-dimensional supergravity
fields in section \ref{sec:ansatz}. Section \ref{sec:no-conical}
describes how the Killing spinor equations reduce to one single second
order differential equation, and uses it to comment on the conditions
required to eliminate conical singularities in the metric. The
following sections describe the explicit solutions that we have
found. They are ordered in increasing complexity of the base $B_6$:
section \ref{sec:KE6} deals with $B_6=KE_6$, section \ref{sec:KE4xKE2}
with $B_6=KE_4\times KE_2$, and section \ref{sec:KE2xKE2xKE2} with
$B_6=KE_2 \times KE_2 \times KE_2$. The type IIB solutions that arise
when one of the $KE_2$ factors is a $T^2$ are discussed in section
\ref{iib}. A discussion of the new non-compact solutions of type IIB
supergravity is presented in section~\ref{sec:noncompact} and we
conclude in section \ref{sec:conclusions}. Finally, we have included
three appendices. The technical study of the Killing spinor equations
has been relegated from the text, and collected in appendix
\ref{sec:appendix}. In particular, this appendix contains the
classification of the relevant $AdS_3$ solutions using
G-structures. Some properties of $KE_4$ spaces used in this paper are
explained in appendix \ref{sec:appendix-KE4}. Finally, appendix
\ref{sec:appendix-orientations} describes how to relate the
orientations and parameters in M-theory and IIB supergravity.


\section{Ansatz and Solutions}
\label{sec:ansatz}


To explain our ansatz in a bit more detail, we recall that the family of $AdS_5$ solutions
of~\cite{GMSW} with $M_6$ an $S^2$ bundle over an $H^2\times S^2$ base-space contain a
limiting solution in which the geometry degenerates to $H^2\times
S^4$. This particular
solution was constructed previously by Maldacena and N\'u\~nez and describes the near horizon limit
of M-fivebranes wrapping a holomorphic two-cycle (calibrated by the K\"ahler two-form)
inside a Calabi-Yau three-fold~\cite{mn}. Here the
two-cycle is the $H^2$ factor\footnote{It is possible to take a quotient of
$H^2$ to obtain any compact Riemann surface with genus greater than one whilst still preserving
supersymmetry~\cite{mn}.} while the $S^4$ factor corresponds to the four-sphere surrounding the wrapped
fivebranes. The metric for this ``2-in-6 K\"ahler'' solution contains
an $S^4$ factor twisted over the $H^2$ base in a specific manner, which can be
deduced from probe fivebranes wrapping holomorphic cycles~\cite{mn}.
From this perspective, the solutions of~\cite{GMSW} correspond to
a generalisation where the $S^4$ surrounding the brane is replaced by an
$S^2$ bundle over $S^2$.
\vskip .3cm
\begin{table}[h]
\label{array}
\begin{center}
\begin{tabular}{l|ccccccccccc}
``2-in-6'' & 0 &1&2&3&4&5&6&7&8&9&10
\cr \hline
 $CY_3$ &  && &  &- &-&-&-&-&-&
 \cr
M5 &-&-&-&-&\multicolumn{2}{c}{$\underbrace{\mbox{\,-\,\,\,\,\,\,  - }}$}
 \cr
\multicolumn{2}{c}{}&&&& \multicolumn{2}{c}{$H_2$}
\end{tabular}
\vskip .5cm
\begin{tabular}{l|ccccccccccc}
``4-in-8'' & 0 &1&2&3&4&5&6&7&8&9&10
\cr \hline
 $CY_4$ &  &&-&-&-& -&- &-&-&-&
 \cr
M5 &-&-& \multicolumn{4}{c}{$\underbrace{\mbox{\,-\,\,\,\,  - \,\,\,  - \,\,\,\,\,  - }}$}
 \cr
\multicolumn{2}{c}{}&& \multicolumn{4}{c}{$KE_4$}
\end{tabular}
\end{center}
\caption{Arrays describing the probe brane setups in M-theory. In both situations, of the 5 transverse
directions to the M5-branes, 4 are tangent to the CY and 1 lies in flat space.}
\end{table}

For us, a key observation is that the structure of the $AdS_5$ 2-in-6 K\"ahler solution~\cite{mn}
has many similarities with the $AdS_3$ solution of~\cite{GKW} corresponding
to M-fivebranes wrapping a K\"ahler four-cycle in a Calabi-Yau
four-fold, ``4-in-8 K\"ahler''.
In both cases, of the five directions transverse to the probe M-fivebranes wrapping such cycles,
four of them are tangent to the Calabi-Yau. Combined with
the fact that in both cases the fivebrane is wrapping a K\"ahler cycle, this implies
that in the $D=11$ solution the four-sphere is fibred over the base space in an almost identical
way, the only difference being that the $H^2$ base is replaced by a
four-dimensional K\"ahler--Einstein manifold $KE_4$ that is negatively
curved. We have summarised the probe-brane configurations in Table
1. Note that in the 4-in-8 case one could also consider a probe
membrane wrapping directions orthogonal to the Calabi-Yau without
further breaking the supersymmetry. 

Thus, inspired by the fact that the 2-in-6 K\"ahler solution is a
single member of a family of solutions where the $S^4$ is replaced by
an $S^2$ bundle over $S^2$, we are motivated to find new warped $AdS_3\times\CM_8$ solutions where
$\CM_8$ is a bundle over $KE_4$, with fibres which are themselves $S^2$
bundles over $S^2$. Or rather, in analogy with~\cite{GMSW}, the structure should
actually simplify to an $S^2$ bundle over $S^2\times KE_4$.
In the specific 4-in-8 K\"ahler solution of~\cite{GKW} the $KE_4$ had
negative curvature . However, following the success of~\cite{GMSW} we
can also seek solutions that are $S^2$ bundles over more general bases
$B_6$. Specifically we will consider $B_6=KE_6$, $B_6=KE_2\times KE_4$
or $B_6=KE_2\times KE_2\times KE_2$, with the various factors having
positive, negative or zero curvature.

\subsection{Ansatz}

We start with the ansatz for the bosonic fields of $D=11$
supergravity, discussing each choice for $B_6$ in turn. In all cases we
assume the metric is a warped product:
\be\label{an1}
ds^2=\w^2 \left[ ds^2(AdS_3)+ ds^2(\CM_8)\right] \,,
\ee
where $ds^2(AdS_3)$ is the metric of constant curvature on $AdS_3$
with unit radius.

\subsubsection*{Case 1: $B_6=KE_2\times KE_2\times KE_2$}

In this case we assume that $ds^2(\CM_8)$ has the form
\be\label{an2}
ds^2(\CM_8)=\sum_{i=1}^3 h_i ds^2(C_i)+ f_3 dr^2 + f_4 (D\psi)^2 \,,
\ee
and the four-form is given by
\bea\label{an3}
G_4 &=& g_3 J_1\wedge J_2 +g_1 J_2\wedge J_3 +g_2 J_3\wedge J_1
\nn && \quad
+ \left( g_4 J_1+g_5 J_2+g_6 J_3 \right)\wedge dr\wedge D\psi +g_7 dr\wedge \Vol{AdS_3} \,.
\eea
Here
$ds^2(C_i)$ is locally a constant curvature metric on $S^2$, $H^2$ or
$T^2$ for $k_i=1,-1$ or $0$, respectively, and $J_i=\Vol{C_i}$ is the
corresponding K\"ahler-form on $C_i$. The ansatz depends on thirteen
functions $h_i$, $f_3,f_4$, $g_1,\dots,g_7, \w$ which are functions of
$r$ only. Thus, in general, the isometry group will be, at least,
$SO(2,2)\times U(1)$, the first factor corresponding to the symmetries
of $AdS_3$ and the latter to shifts of the fiber coordinate $\psi$ (we
will see later that, for compact $\CM_7$, $\psi$ is a periodic
coordinate). Note that using the freedom to change the $r$ coordinate
we can choose $f_3$ as we like. We also have
\be
D\psi=d\psi+P
\ee
with
\be
dP= \sum_{i=1}^3{\cal R}_i=\sum_{i=1}^3 k_i J_i \,,
\ee
where ${\cal R}_i$ is the Ricci-form for $C_i$.
Thus, the twisting of the fibre with coordinate $\psi$
is associated to the canonical $U(1)$ bundle over the six-dimensional
base space $B_6$ given by $C_1\times C_2\times C_3$. Indeed in the
compact complete solutions that we will construct $r,\psi$ will
parametrise a two-sphere and topologically we will have an $S^2$
bundle over $C_1\times C_2\times C_3$ that is obtained by adding a
point to the fibres in the canonical line bundle. This is entirely
analogous to the solutions in~\cite{GMSW}.

\subsubsection*{Case 2: $B=KE_4\times KE_2$}

In this case we assume
\bea
ds^2(\CM_8)&=&h_1 ds^2(KE_4)+ h_3 ds^2(C_3)+f_3 dr^2 + f_4 (D\psi)^2
\,, \nn
G_4 &=& \frac{g_3}{2} J\wedge J +g_1 J\wedge J_3 \nn
 &&\quad  {}+ \left( g_4 J+g_6 J_3 \right)\wedge dr\wedge D\psi
 + g_7 dr\wedge \Vol{AdS_3} \,,
\eea
where now $J$ is the K\"ahler form of $KE_4$. Note that in terms of
the functions $h_i$, $g_i$ and parameters $k_i$, this can be viewed as
a special case of the previous ansatz
\be\label{KE4}
k_1=k_2 \sac h_1=h_2 \sac g_1=g_2 \sac g_4=g_5 \,.
\ee
The 4-in-8 K\"ahler solution of~\cite{GKW} is contained within this ansatz.
More specifically, we shall show that it is recovered in the
$B_6=KE_4\times KE_2$ class \bref{KE4} when $KE_4$ has negative curvature ($k_1=k_2=-1$)
and $C_3=S^2$ ($k_3=1$). For this particular solution, the $r,\psi$ coordinates parametrise a two-sphere
but when combined with the two sphere $C_3$ give rise to a four-sphere. The metric for this solution is
then a warped product of the form $AdS_3\times KE_4\times S^4$. From a physical point of view
the four-sphere surrounds the fivebrane that is wrapped on the negatively
curved\footnote{Note that solutions for fivebranes wrapping positively
  curved K\"ahler-Einstein spaces, i.e. $k_2=1$ were also found
  in~\cite{GKW} but they did not have an $AdS_3$ factor in the
  near-brane limit.} $KE_4$.

\subsubsection*{Case 3: $B_6=KE_6$}

In this case the ansatz is even simpler
\bea
ds^2(\CM_8)&=&h_1 ds^2(KE_6)+ f_3 dr^2 + f_4 (D\psi)^2
\,, \nn
G_4 &=& \frac{g_1}{2}J\wedge J +  g_4 J\wedge dr\wedge D\psi +g_7 dr\wedge \Vol{AdS_3}
\,,
\eea
where here $J$ is the K\"ahler form of $KE_6$. Again in terms of the
functions and parameters, this is a special case of the original
ansatz with
\be
\label{KE6-conditions}
k_1=k_2=k_3 \sac h_1=h_2=h_3 \sac g_1=g_2=g_3 \sac g_4=g_5=g_6 \,.
\ee

\bigskip

Finally, our ansatz leads to type IIB backgrounds when one of the
cycles in $B_6$ is a torus. By reducing to IIA along one of the torus
directions and T-dualising along the other, we obtain type IIB
solutions with constant dilaton and only 5-form flux. To be explicit,
let $C_3=T^2$. The eleven-dimensional four-form decomposes naturally
as
\bea
G_4 &=& F_4 + F_2 \wedge \Vol{T^2}
\,,\nn
F_4 &=& g_3 J_1 \wedge J_2 + (g_4 J_1+g_5 J_2)\wedge \, dr \wedge
D\psi + g_7 dr \wedge \Vol{AdS_3} \,,\nn
F_2 &=& g_2 J_1+g_1 J_2 + g_6 dr \wedge D\psi \,.
\eea
Note that both $F_2$ and $F_4$ are closed by virtue of the closure of
$G_4$. In particular, we can write
locally $F_2=dA_1$ for some one-form potential $A_1$. Having made this
decomposition, the IIB background reads (for a few more details see
appendix \ref{sec:appendix-orientations})
\bea
ds^2&=& (\w^6 h_3)^{1/2} \left[ ds^2(AdS_3) + ds^2(\CM_7) \right] \,, \nn
F_5 &=& (1+ \star) F_4 \wedge Dz \,,
\label{general-IIB}
\eea
where
\begin{equation}
\label{IIBmetric}
   ds^2(\CM_7) = h_1 ds^2(C_1) +h_2 ds^2(C_2)
     + f_3 dr^2+f_4 D\psi^2 +\frac{1}{\w^6h_3^2 }Dz^2 \,,
\end{equation}
and
\bea
Dz=dz+A_1 \,.
\eea
For our ansatz, explicitly we have
\begin{equation}
\begin{aligned}
   \star\left[F_4\wedge Dz\right] &=
   \frac{h_3g_3\w^3 \sqrt{f_3f_4}}{h_1h_2} \Vol{AdS}\wedge dr \wedge D\psi
   + \frac{h_2 h_3 g_4 \w^3}{h_1\sqrt{f_3f_4}} \Vol{AdS_3}\wedge J_2 \\
   & \quad
   {}+ \frac{h_1 h_3 g_5 \w^3}{h_2\sqrt{f_3f_4}} \Vol{AdS_3}\wedge J_1
   + g_7 h_1h_2 h_3\w^3 \sqrt{\frac{f_4}{f_3}}
      J_1 \wedge J_2 \wedge D\psi \,.
\end{aligned}
\end{equation}
Note that (as explained in appendix~\ref{sec:appendix}), taking
$\w^3>0$, we have $\sqrt{f_3 f_4} <0$. Also note that these IIB
backgrounds will have an isometry group at least as large as
$SO(2,2)\times U(1) \times U(1)$, the $U(1)$ factors acting as shifts
of the coordinates $\psi,z$.


\subsection{The BPS condition and global construction}
\label{sec:no-conical}


Let us now turn to the conditions imposed by supersymmetry, the
Bianchi identities and the equations of motion. The derivation is
technically involved and relies on a key gauge choice for the function
$f_3$. We have presented some details in
appendix~\ref{sec:appendix}. Remarkably, we find that all conditions boil down to solving the second-order
non-linear equation for a function $H(r)$:
\begin{equation}
\label{Heq}
\begin{aligned}
   0 &= -4 (H')^2 + 4 H\left(
      2 H''+4 k_i d_i + 4  k_i c_i r -3 k_1k_2k_3r^2 \right) \\
      & \qquad {}
      + \prod_{i=1}^3 \left({k_1k_2k_3 \over k_i}r^2-4 r c_i -4d_i \right)
      \,.
\end{aligned}
\end{equation}
Here $c_i$ and $d_i$ are six integration constants that appear in the
analysis. Given a solution of~\eqref{Heq} one can construct the full
eleven-dimensional supergravity solution as follows
\bea
h_l&=& {2 H''+4 k_i d_i + 4 k_i c_i r -3 k_1 k_2 k_3 r^2 \over 4 (4 d_l+4 c_l r - {k_1 k_2 k_3\over k_l} r^2) }
\sac l=1,2,3 \,,
\nn
f_3 &=& -{ 2 H''+4 k_i d_i + 4 k_i c_i r -3 k_1 k_2 k_3 r^2 \over 16 H}
\,,\nn
f_4&=& -H \,\, { 2 H''+4 k_i d_i + 4 k_i c_i r -3 k_1 k_2 k_3 r^2
\over
 \prod_{i=1}^3 ({k_1k_2k_3 \over k_i} r^2 -4 r c_i -4d_i ) } \,,
\nn
\w^6 &=& {4 \prod_{i=1}^3 ({k_1k_2k_3 \over k_i} r^2 -4 r c_i -4d_i ) \over
\left( 2 H''+4 k_i d_i + 4  k_i c_i r -3 k_1k_2k_3 r^2 \right)^2} \,.
\label{recons}
\eea
If we define the function $f$ via
\bea
f&=& {4 H' \over 2 H''+4 k_i d_i + 4  k_i c_i r -3 k_1k_2k_3 r^2} \,,
\eea
then the first three components of the flux are given by
\bea
g_1 &=&-\undos \left[ f (k_2 h_3 + k_3 h_2) + k_2 k_3 r \right] + c_1\,,
\nn
g_2 &=&-\undos \left[ f (k_3 h_1 + k_1 h_3) + k_3 k_1 r \right] + c_2 \,,
\nn
g_3 &=&-\undos \left[ f (k_1 h_2 + k_2 h_1) + k_1 k_2 r \right] + c_3\,.
\label{bianchis}
\eea
The next three components are given by
\bea
-4 g_4 &=&(2f h_1)'+k_1,\nn
-4 g_5 &=&(2f h_2)'+k_2,\nn
-4 g_6 &=&(2f h_3)'+k_3,
\eea
while the electric piece is
\bea
g_7&=& {f_3 \over 2 h_1h_2h_3} \Big[ (\w^6+f^2) (k_1 h_2 h_3 +\mbox{perm.} )
+ f(k_1 k_2 h_3+\mbox{perm.})r -2 fc_i h_i\Big]
\,, \nonumber
\eea
where ``${}+\mbox{perm.}$'' means adding the other two terms involving
the obvious permutations of $1,2$ and $3$.
The explicit expression for the Killing spinors are given in the Appendix.
We find that the solution generically preserves 1/8 supersymmetry and
is dual to a two-dimensional conformal field theory with $N=(0,2)$
supersymmetry.

We have not managed to find the most general solution to the
differential equation~\eqref{Heq} for $H$. However, we have found a rich set
of polynomial solutions that lead to regular and compact
solutions, which we will discuss in detail in the next section.

Our principal interest is the construction of compact solutions. As
mentioned earlier, our procedure will be to require $r,\psi$ to
parametrise a two-sphere fibred over a compact base $B_6$:
\begin{equation}
\label{fibration}
   \begin{CD}
      S^2_{r,\psi} @>>> \mathcal{M}_8 \\
      && @V{\pi}VV \\
      && B_6
   \end{CD}
\end{equation}
This is achieved if the range of $r$ is restricted to lie in a
suitable finite interval, $r_1\le r\le r_2$, and $\psi$ is a periodic
coordinate. Clearly, the range of the coordinate $r$ is restricted by
the poles of $f_3$ which must be zeroes of the function
$H$. Generically, the poles of $f_3$ are also the zeroes of $f_4$ and
so the $(r,\psi)$ part of the metric can indeed form an $S^2$ provided
that one can remove potential singularities at each of the zeroes
$r_\a$ of $H$. If such a generic $H$ has a linear behaviour at $r_\a$
we find that, after a change of coordinates $r=r(\rho)$, the
$(r,\psi)$ part of the metric takes the form
\bea
ds^2 = d\rho^2 + \g \rho^2 d\psi^2 \,,
\eea
with
\bea
\g &=& {4 [H'(r_\a)]^2 \over \prod_{i=1}^3 ({k_1k_2k_3 \over k_i} r^2_\a -4 r_\a c_i -4d_i )} \,,
\eea
where $H'(r_\a)$ is the value of $H'$ at the corresponding zero. Now, using the differential
equation \bref{Heq}, and evaluating it at a zero of $H$, one deduces the remarkable
fact that $\g=1$ at all poles. Thus, for such generic $H$, the condition for the absence of conical singularities at the poles of the $S^2$
formed by $(r,\psi)$ is just $\Delta \psi= 2\pi$.

For such $H$, we have complete metrics on $\CM_8$, then, provided that
the warp factor $\w$ and the metric functions $h_i$ remain finite and
non-zero within the interval $r_1\le r\le r_2$. Topologically, $\CM_8$
is a two-sphere bundle over a six-dimensional base, $B_6$, where
$B_6=KE_6$, $KE_4\times KE_2$ or $KE_2\times KE_2\times KE_2$. This
$S^2$-bundle is obtained from the canonical line-bundle over $B_6$ by
simply adding a point ``at infinity'' to each of the fibres.


\subsection{Flux quantization}
\label{flux}

Given a regular compact solution, the final condition that we have a
\emph{bona fide} solution of M-theory is the quantization of the
four-form flux $G_4$. In our ansatz~\eqref{an1} we took $AdS_3$ to
have unit radius, thus we should first reinstate dimensions by
rescaling the metric and background by powers of $L$, the actual radius
of the $AdS_3$ space,
\bea
d\tilde{s}_{11}^2=L^2 ds_{11}^2 \sac \tilde{G}_4=L^3 G_4 \,.
\eea
In our conventions, the quantisation condition~\cite{wit} for the
four-form is that we have integer periods
\begin{equation}
   N(\mathcal{D}) = \int_{\mathcal{D}}  \left[
      \frac{1}{(2\pi l_P)^3}\tilde{G}_4
      - \frac{1}{4}p_1(M_{11}) \right]
      \in \bbZ \,,
\end{equation}
where $l_P$ is the eleven-dimensional Planck length (see
appendix~\ref{sec:appendix-orientations}), $\mathcal{D}$ is any
four-cycle of the eleven-dimensional manifold $M_{11}$ and $p_1(M_{11})$
is the first Pontryagin class of $M_{11}$. The subtlety
here is the shift by $\frac{1}{4}p_1(M_{11})$ which is not necessarily
an integral class.

For our solutions $M_{11}=AdS_3\times\CM_8$, and the relevant four-cycles lie
in $\mathcal{M}_8$. Thus, written in terms of cohomology the condition
becomes
\begin{equation}
\label{quant}
   \frac{1}{(2\pi l_P)^3}[\tilde{G}_4]
      - \frac{1}{4}p_1(\mathcal{M}_8) \in H^4(\mathcal{M}_8,\bbZ)
\end{equation}
where $[\tilde{G}_4]$ denotes the cohomology class of $\tilde{G}_4$ on
$\mathcal{M}_8$. Given the fibration structure~\eqref{fibration} it is
relatively easy to find an expression for $p_1(\mathcal{M}_8)$ in terms
of classes on $B_6$. One way to view the fibration~\eqref{fibration}
is as a complex space formed by the anti-canonical line bundle
$\mathcal{L}$ of $B_6$ together with a point $r=r_1$ added at infinity
of each $\bbC$ fibre to make them into spheres. One finds
\begin{equation}
\begin{aligned}
   p_1(\mathcal{M}_8) &= \pi^*p_1(B_6) + \pi^*p_1(\mathcal{L}) \\
      &= \pi^*\left[c_1(B_6)^2 - 2c_2(B_6) \right]
         + \pi^*\left[c_1(B_6)^2\right]
\end{aligned}
\end{equation}
where the second term in the first line comes from the twisting of the
fibration. In the second line we have used the complex structures
on $TB_6$ and $\mathcal{L}$ to rewrite the Pontryagin classes in terms of
Chern classes.

As we will see in the next section all our new solutions will be of
the form
\begin{equation}
   B_6 = B_4 \times C
\end{equation}
where $C$ is some Riemann surface. Given the projections
$\pi_1:\mathcal{M}_8\to B_4$ and $\pi_2:\mathcal{M}_8\to C$ we then
have $\pi^*c_1(B_6)=\pi_1^*c_1(B_4)+\pi_2^*c_1(C)$ and
$\pi^*p_1(B_6)=\pi_1^*p_1(B_4)+\pi_2^*p_1(C)=\pi_1^*p_1(B_4)$. Hence
\begin{equation}
\label{ddd}
\begin{aligned}
   p_1(\mathcal{M}_8) &=
      \pi_1^*\left[ 2c_1(B_4)^2 - 2c_2(B_4) \right]
         + 2\pi_1^*c_1(B_4)\wedge\pi_2^*c_1(C) , \\
      &= \pi_1^*\left[ 4c_1(B_4)^2 - 24\chi(\mathcal{O}_{B_4})\right]
         + 4\pi_1^*c_1(B_4)\wedge\pi_2^*\chi(\mathcal{O}_C) . \\
\end{aligned}
\end{equation}
where the alternating sum of Hodge numbers
$\chi(\mathcal{O}_M)=\sum_p(-1)^ph^{0,p}(M)$ is the holomorphic Euler
characteristic. (For $B_4$~\cite{GH,Beauville} we have
$\chi(\mathcal{O}_{B_4})=\frac{1}{12}[c_1(B_4)^2+c_2(B_4)]$, while for
a Riemann surface it is half the Euler number
$\chi(\mathcal{O}_C)=\frac12\chi=1-g$, where $g$ is the genus.)

From~\eqref{ddd} we see that $\frac{1}{4}p_1(\mathcal{M}_8)$ is in
fact integral for our new examples. Thus in the following we may
neglect the $\frac{1}{4}p_1(\mathcal{M}_8)$ term in~\eqref{quant} and
simply require that $[\tilde{G}_4]/(2\pi l_P)^3$ is integral.


\section{The class $B_6=KE_6$}
\label{sec:KE6}


Let us begin with the simplest case: $B_6=KE_6$. As explained above, we need
to impose the conditions \bref{KE6-conditions}, which imply
$c_3=c_2=c_1$ and $d_3=d_2=d_1$. We are therefore left with only two
integration constants, say $c_1,d_1$, and also the constant $k_1$
specifying the curvature of the $KE_6$. We discuss the cases $k_1 \neq
0$  and $k_1=0$ separately.

\subsection{$B_6=KE_6^{\pm}$}

From \bref{bianchis} we see that when $k_1 \neq 0$ we can shift the
coordinate $r$ to set $c_1=0$. Having done this, we are left with only
one independent constant, say $d_1$. Our polynomial solution for $H$
is
\be
H=k_1 \left( \frac{1}{4} r^4 - 2 d_1 r^2+4 d_1^2 \right) \,,
\ee
which leads to the following metric
\bea
ds^2&=& \w^2 \left[ ds^2(AdS_3) -  {4d_1+3r^2 \over 4(r^2-4d_1)}
  \left( k_1 ds^2(KE_6)+D\psi^2 + {dr^2 \over r^2 - 4d_1} \right)\right]
\nonumber\,,
\eea
with
\bea
\w^6 &=& {4(r^2-4d_1)^3 \over (3r^2+4d_1)^2} \,.
\eea
Demanding that the metric is positive definite implies that we must
take $k_1=1$, choose $d_1<0$ and restrict the range of $r$ so that
$r^2 \le 4|d_1|/3$. However, at $r^2=4|d_1|/3$, the warp factor $\w$
diverges and hence there are no compact regular solutions in this
class.

\subsection{$B_6=CY_3$}

If $k_1=0$ we have $B_6=CY_3$ (this includes $B_6=T^6$ and $B_6=CY_2\times T^2$). We have only found
one single regular solution, for which
\be
H= -4 r^2+1 \,,
\ee
and $c_i=0, d_i=-1$. This leads to constant metric functions:
\bea
h_1=h_2=h_3=\w=1 \,.
\eea
A redefinition $y=2r$ then yields
\bea
ds^2&=&ds^2(AdS_3) + ds^2(CY_3) +
\frac{1}{4}\left[\frac{dy^2}{1-y^2} +(1-y^2)d\psi^2\right]
\,,\nn
G_4 &=& \frac{1}{2}J\wedge \Vol{S^2} \,,
\label{ads3cy3}
\eea
where $\Vol{S^2}=dy\wedge d\psi$ is the volume form of the round $S^2$
parametrised by $y,\psi$. This solution is thus simply the well known
$AdS_3\times S^2\times CY_3$ solution that is dual to a $(0,4)$
superconformal field theory. In the special case that $CY_3$ is
$CY_2\times T^2$ we can dimensionally reduce and T-dualise to obtain
the type IIB solution of the form $AdS_3\times CY_2\times S^3$ with
non-zero five-form flux, which is the near horizon limit of two
intersecting D3-branes and is dual to a $(4,4)$ superconformal field
theory.


\section{The class $B_6=KE_4\times KE_2$}
\label{sec:KE4xKE2}


The next to simplest case is when the base is $B_6=KE_4 \times KE_2$,
which happens when we impose the conditions \bref{KE4}. These imply
that we have to set $c_2 = c_1$ and $d_2 = d_1$, leaving a total of
four independent constants, $\{c_1,c_3,d_1,d_3 \}$ say, and the two
curvature constants $k_1$ and $k_3$ of $KE_4$ and $KE_2$,
respectively. We proceed to consider the cases $k_1 =0$ and $k_1 \neq
0$ separately.


The case $k_1=0$, which leads to $B_6=K3 \times KE_2$ or $B_6=T^4 \times KE_2$,
is rather special, as we cannot set to zero any of the $c_i$ by shifting $r$.
Thus the general solution is specified by $\{c_1,c_3,d_1,d_3\}$ and the $KE_2$ curvature $k_3$.
We have found the following cubic polynomial solution for $H$:
\bea
H&=&-\frac{4 \text{c_3} }{3 \text{k_3}}r^3 \, -\frac{4 \left(3 \text{c_1}^2+\text{d_3}\right) }{\text{k_3}}r^2
    -\frac{4 \left(6 \text{c_1} \text{d_1} \text{c_3}^2+3 \text{d_3}
    \left(4 \text{c_1}^2+\text{d_3}\right) \text{c_3}\right) }{3 \text{c_3}^2 \text{k_3}} r
\nn&&
   -\frac{4 \left(\text{c_3}^2 \text{d_1}^2+4 \text{c_1} \text{c_3} \text{d_3}
   \text{d_1}+\text{d_3}^2 \left(4 \text{c_1}^2+\text{d_3}\right)\right)}{3 \text{c_3}^2 \text{k_3}} \,,
\eea
which leads to the metric functions
\bea
h_1&=& -\frac{3 \left(4 \text{c_1}^2+\text{d_3}+\text{c_3} r\right)}{4 \text{k_3} (\text{d_1}+\text{c_1} r)}
\,,\nn
h_3&=& -\frac{3 \left(4 \text{c_1}^2+\text{d_3}+\text{c_3} r\right)}{4 \text{k_3} (\text{d_3}+\text{c_3} r)}
\,, \nn
\w^6&=& -\frac{16 (\text{d_1}+\text{c_1} r)^2 (\text{d_3}+\text{c_3} r)}{9 \left(4 \text{c_1}^2+\text{d_3}+\text{c_3} r\right)^2}
\,, \nn
f_3&=&\frac{3 \left(4 \text{c_1}^2+\text{d_3}+\text{c_3} r\right)}{\text{4 H k_3}}
\,, \nn
f_4&=&-\frac{3 H \left(4 \text{c_1}^2+\text{d_3}+\text{c_3} r\right)}{16 \text{k_3} (\text{d_1}+\text{c_1} r)^2 (\text{d_3}+\text{c_3} r)}
\,.
\eea
Unfortunately, we have not been able to prove the existence of positive definite metrics leading to compact solutions
for any choice of $\{c_1,c_3,d_1,d_3,k_3\}$.


However, if $k_1 \neq 0$, then we can set $c_3=0$ by shifting $r$.
Thus the general solution is specified by $\{c_1,d_1,d_3\}$ and the
$KE_4$ and $KE_2$ curvatures $k_1$ and $k_3$. We have found the
following quartic polynomial solution for $H$:
\be
H= \sum_{n=0}^4 p_n r^n \,,
\ee
where
\bea
p_4 &=& \tfrac{1}{4}k_1^2 k_3 \,,
\nn
p_3 &=& - \tfrac{2}{3} k_1 c_1\,,
\nn
p_2&=& \frac{3 \text{d_1}^2 \text{k_1}^2}{2 \text{d_1} \text{k_1}+\text{d_3} \text{k_3}}-2 \text{d_1} \text{k_1}-\text{d_3} \text{k_3}\,,
\nn
p_1&=& \frac{8 \text{c_1} \text{d_1} \text{d_3}}{2 \text{d_1} \text{k_1}+\text{d_3} \text{k_3}} \,,
\nn
p_0&=& \frac{4 \text{d_3} \left(4 \text{d_3} \text{c_1}^2+(2 \text{d_1} \text{k_1}+\text{d_3} \text{k_3})^2\right)}{3 \text{k_1}^2 (2 \text{d_1} \text{k_1}+\text{d_3} \text{k_3})}
\,.
\label{KE4-H}
\eea
The corresponding expressions for the metric functions are:
\bea
h_1&=& \frac{3 \text{k_1}^2 \left(4 \text{d_1}^2+\text{k_3} (2 \text{d_1} \text{k_1}+\text{d_3} \text{k_3}) r^2\right)}{4 (2 \text{d_1} \text{k_1}+\text{d_3} \text{k_3}) (4
   \text{d_1}+r (4 \text{c_1}-\text{k_1} \text{k_3} r))}
\,, \nn
h_3&=& -\frac{3 \text{k_1}^2 \left(4 \text{d_1}^2+\text{k_3} (2 \text{d_1} \text{k_1}+\text{d_3} \text{k_3}) r^2\right)}{4 (2 \text{d_1} \text{k_1}+\text{d_3} \text{k_3})
   \left(\text{k_1}^2 r^2-4 \text{d_3}\right)}
\,, \nn
\w^6&=&\frac{4 (2 \text{d_1} \text{k_1}+\text{d_3} \text{k_3})^2 \left(\text{k_1}^2 r^2-4 \text{d_3}\right) (4 \text{d_1}+r (4 \text{c_1}-\text{k_1} \text{k_3} r))^2}{9 \text{k_1}^4
   \left(4 \text{d_1}^2+\text{k_3} (2 \text{d_1} \text{k_1}+\text{d_3} \text{k_3}) r^2\right)^2}
\,, \nn
f_3 &=&-\frac{3 \text{k_1}^2 \left(4 \text{d_1}^2+\text{k_3} (2 \text{d_1} \text{k_1}+\text{d_3} \text{k_3}) r^2\right)}{16 \text{H} (2 \text{d_1} \text{k_1}+\text{d_3}
   \text{k_3})}
\,, \nn
f_4 &=& \frac{3 \text{H} \text{k_1}^2 \left(4 \text{d_1}^2+\text{k_3} (2 \text{d_1} \text{k_1}+\text{d_3} \text{k_3}) r^2\right)}{(2 \text{d_1} \text{k_1}+\text{d_3} \text{k_3})
   \left(4 \text{d_3}-\text{k_1}^2 r^2\right) (4 \text{d_1}+r (4 \text{c_1}-\text{k_1} \text{k_3} r))^2}
\,.
\label{KE4-metrics}
\eea
We have found positive definite metrics that lead to compact solutions in the classes where $k_1 k_3=-1$,
i.e. $KE_4^-\times S^2$ and $KE_4^+\times H^2$, and $k_3=0$, i.e. $KE_4^+\times T^2$. We now
proceed to discuss them separately.

\subsection{$B_6=KE_4^- \times S^2$ and the solution of~\cite{GKW}}
\label{sec:GKW-class}

Let $(k_1,k_3)=(-1,1)$. In this section we argue that there exists a range of the constants $\{c_1,d_1,d_3\}$
for which the expressions \bref{KE4-metrics} yield  positive definite metrics and compact $\CM_8$.
To show this we first note that the special point in the $\{c_1,d_1,d_3\}$ space defined by
\bea
\label{gkw-constraint}
c_1 = 0 \sac d_1 = -9/16 \sac d_3 =-27/16  \,,
\eea
is actually the 4-in-8 K\"ahler solution of~\cite{GKW} discussed in section \ref{sec:ansatz}. Indeed, defining $y=2r/3$ we have
\bea
\w^6 &=& 3+y^2 \sac h_1={3 \over 4} \sac h_3 =f_4 = {3 (1-y^2) \over 4 (3+y^2)} \sac  f_3= {3 \over 4(1-y^2)} \nonumber\,.
\eea
A further change $y=\cos \a$ leads to the 4-in-8 K\"ahler solution of
\cite{GKW}:
\bea\label{coordchgkw}
ds^2(\CM_8)= {3 \over 4} \left[ ds^2(KE_4^-)+d\a^2+{\sin^2\a \over
3+\cos^2\a} \left[ds^2(S^2)+D\psi^2\right] \right]  \,.
\eea
If the range of $\psi$ is taken to be $4\pi$ then, at fixed $\alpha$
the $S^2$ and $\psi$ parametrise a round three-sphere and together
with $\alpha$ parametrise a four-sphere. Clearly, this solution is regular
and compact (provided $KE_4^-$ is). We can recover
the solution in the coordinates used in~\cite{GKW} by defining constrained
coordinates $Y^A$ on the $S^4$,
satisfying $Y^A Y^A=1$, via
\bea
Y^1+iY^2&=&\sin\alpha\cos\frac{\theta}{2}e^{\frac{i}{2}(\nu-\psi)}\nn
Y^3+iY^4&=&\sin\alpha\sin\frac{\theta}{2}e^{-\frac{i}{2}(\nu+\psi)}\nn
Y^5&=&\cos\alpha
\eea
where $\theta,\nu$ are polar coordinates for the $S^2$ appearing
in~\eqref{coordchgkw}.

It can be easily checked that for values of the constants sufficiently
close to~\bref{gkw-constraint}, the metric is still regular and
positive definite for values of $r$ between the two relevant roots of
$H$. The topology of these nearby solutions is however not
$KE_4^-\times S^4$. The function $h_3$ now does not go to
zero. Instead, by virtue of the general properties discussed in
section~\ref{sec:no-conical}, taking the period of $\psi$ to be
$2\pi$, the $\psi$-circle fibers over $r$ form a smooth
$S^2$. Globally the manifold is constructed by fibering this $S^2$
over $B_6=KE_4^-\times S^2$. Noting that we can still use the
scaling symmetry of the eleven-dimensional equations of motion to set
either $d_1$ or $d_3$ to a fixed value, we conclude that the
expressions~\bref{KE4-metrics} (with $k_1=-k_3=-1$) lead to a
two-parameter family of solutions that include the solution
of~\cite{GKW} as a special case in which the topology changes. This
family of solutions thus realises expectations of
section~\ref{sec:ansatz}.


\subsection{$B_6=KE_4^+ \times H^2$}
\label{sec:inverse-GKW-class}


We repeat the analysis of the previous section in the
$(k_1,k_3)=(1,-1)$ class. A very simple regular and compact solution
occurs at the point
\bea
\label{toni-constraint}
c_1 = 0 \sac d_1 = d_3 = -1/4  \,,
\eea
where the metric is
\bea
ds^2&=&\w^2 \left[ ds^2(AdS_3)+{3(1+r^2)\over 4(1-r^2)} ds^2(KE_4^+)
   + {3 \over 4} ds^2(H^2) \right.
   \nn && \left.\,\,\,\,\,\,\,\,\,\,\,\,
   {} +{9(1+r^2) \over 4 q(r)} dr^2
   + {q(r) \over 4(1-r^2)^2} D\psi^2 \right] \,,
\label{new}
\eea
with
\be
\w^6 = {4 (1-r^2)^2 \over 9(1+r^2)} \sac q(r) = 1-6r^2-3r^4 \,.
\ee
It can be readily checked that the quartic polynomial $q(r)$ has only
two real roots of opposite sign $r_2=-r_1$, with absolute value less
than one, and that $q(r)$ is positive between these two. Although the
analysis of section~\ref{sec:no-conical} does not apply because the
expansion near the roots is not linear, one can still show that if the
range of $r$ is restricted to lie between them, the metric is positive
definite, and $r_1,r_2$ are the north/south poles of the $S^2$ formed
by $(r,\psi)$. Note that one can easily obtain compact solutions by
the standard procedure of taking the quotient of $H^2$ by a discrete
element of $SL(2,\bbZ)$, to obtain a Riemann surface with genus
greater than one. As shown in appendix~\ref{app:KS}, the Killing
spinors are independent of the coordinates on $H^2$, and are therefore
preserved in the quotient procedure.

Having concluded that this solution is regular and compact, one can
now check that for values of the constants sufficiently close to
\bref{toni-constraint}, the metric is still regular and positive
definite between the two relevant roots of $H$. By virtue of the
general properties discussed in section~\ref{sec:no-conical}, the
$\psi$-circle always fibers over an interval parametrised
by $r$ to form a smooth $S^2$. Thus, noting that we can still use the
scaling symmetry of the eleven-dimensional equations of motion to set
either $d_1$ or $d_3$ to a fixed value, we conclude that the
expressions \bref{KE4-metrics} (with $k_1=-k_3=1$) lead to a
two-parameter family of deformations of the solution~\bref{new}.


\subsection{$B_6=KE^+_4 \times T^2$}
\label{sec:KE4xT2-class}


To study this class, we just need to set $k_3=0$ in the $KE_4\times
KE_2$ expressions~\bref{KE4-H}--\bref{KE4-metrics}. It can then be seen
that we cannot have $d_1=0$, so it can be set to $\pm 1$ by rescaling
$r$. Further inspection reveals that we need to set $k_1 =1$, $d_1=-1$
in order to have a metric with the right signature and so we are led to
consider only $KE_4^+$ spaces.

We will consider the cases when $c_1=0$ and when $c_1\ne 0$
separately. The solutions in this section are of particular interest
because one can use dimensional reduction and T-duality to relate them
to new type IIB solutions. This will be discussed in section~\ref{iib}
below.

\subsubsection{$c_1=0$ Solutions}
\label{sec:Naka-class}

Setting $c_1=0$, the metric and the warp factor read
\bea
\label{c1=0}
ds^2(\CM_8)&=&\frac{3}{8}ds^2(KE_4^+)+\frac{3}{2(r^2-4d_3)}ds^2(T^2)+\frac{9dr^2}{4(3r^2-16d_3)}\nn&&+\frac{(3r^2-16d_3)D\psi^2}{16(r^2-4d_3)}
\,,\nn
\w^6&=&\frac{16(r^2-4d_3)}{9}
\,.
\eea
A further rescaling allows us to choose $d_3=0,\pm1$. When $d_3=0$ the solution is singular. However, when $d_3=\pm 1$ the
solution is regular. For example, when $d_3=1$, the coordinate change $r=(4/{\sqrt 3})\cosh\alpha$ leads to
the manifestly regular metric and warp factor
\bea
ds^2(\CM_8)&=&\frac{3}{8}ds^2(KE_4^+)+\frac{9}{8(4\cosh^2\alpha -3)}ds^2(T^2)+\frac{3d\alpha^2}{4}+\frac{3\sinh^2\alpha D\psi^2}{4(4 \cosh^2\alpha-3)}\nn
\w^6&=&\frac{64(4\cosh^2\alpha - 3)}{27}
\,.\nonumber
\eea

\subsubsection{$c_1\ne0$ Solutions}
\label{sec:IIB-class}

When $c_1\ne 0$ we can rescale it to $1$. Defining
\be
y=1-r \sac a=1-4d_3 \,,
\ee
the eleven-dimensional background reads
\bea
ds^2 &=& \w^2 \left[ ds^2(AdS_3) + ds^2(\CM_8) \right]
\,,\nn
G_4&=& F_4 + F_2 \wedge \Vol{T^2} \,,
\eea
with
\bea
\label{c1neq0}
\w^6&=&\tfrac{16}{9}\, y^2(y^2-2y+a)
\,,\nn
\tfrac{8}{3}\, ds^2(\CM_8)&=&{1 \over y} ds^2(KE_4^+)
   + {4 \,ds^2(T^2) \over (y^2-2y+a)}  + {6 dy^2 \over q(y)}
   + {q(y) D\psi^2 \over 6y^2(y^2-2y+a)}
      \,,\nn
F_4&=& {y-a \over 8y} J\wedge J + {a \over 8 y^2} J\wedge dy\wedge D\psi - {8y \over 3} dy \wedge \Vol{AdS_3}
\,,\nn
F_2&=& {a-y \over 2(y^2-2y+a)} J+{y^2-2a y+ a \over 2(y^2-2y+a)^2} dy\wedge D\psi
\label{G4-decomp}
\eea
and where $q(y)$ is the cubic polynomial
\be
q(y)=4 y^3 -9 y^2 +6 a y-a^2 \,.
\ee
Again one can show that this leads to a family of regular
solutions parametrised by $a$. We will discuss this in more detail in
the IIB dual formulation in section~\ref{iib}.

\section{The class $B_6=KE_2\times KE_2\times KE_2$}
\label{sec:KE2xKE2xKE2}

In this section we consider the base space to be a product $B_6=KE_2 \times KE_2 \times KE_2$ space. This
case is the most general in the sense that it is associated with the
most general polynomial solution of the differential equation \bref{Heq}.
We have found that the integration constants $\{c_i,d_i\}$ have to be constrained
in order for polynomial solutions to exist. This constraint reads
\bea \label{constraint-general}
0&=& d_1 (k_1^2 d_1+4 c_2 c_3) (k_2 c_2-k_3 c_3) + 2 k_1 d_1 (k_2^2 c_2 d_2-k_3^2 c_3 d_3) + \mbox{perm} \,.
\eea
The solution for $H$ is then a quartic polynomial whose coefficients are rather
long expressions involving the constants $\{c_i,d_i\}$ and the
curvatures $k_i$. We know however that these expressions must be
symmetric under permutations of the indices $\{1,2,3\}$. Indeed, the
only symmetric combinations appearing in $H$ are
\begin{equation}
\begin{gathered}
   k = k_1k_2k_3 \,, \\
   E_c = k_i c_i \,, \qquad
   E_d = k_i d_i \,, \\
   E_{cc} = k_1 k_2 c_1 c_2 + \mbox{perm} \,, \qquad
   E_{dd} = k_1k_2 d_1 d_2 + \mbox{perm} \,, \\
   E_{cd} = k_1 c_1 (k_2 d_2+k_3 d_3) + \mbox{perm} \,, \\
   E_{ccc} = c_1 c_2 c_3 \,, \qquad
   E_{ccd} = c_1 c_2 d_3 + \mbox{perm} \,, \qquad
   E_{ddc} = d_1 d_2 c_3 + \mbox{perm} \,.
\end{gathered}
\end{equation}
Using these definitions $H$ reads
\bea
H &=& \sum_{n=0}^4 p_n r^n \,,
\eea
with
\begin{equation}
\label{generalH}
\begin{aligned}
   p_4 &= \tfrac{1}{4}k \,, \\
   p_3 &= -\tfrac{2}{3} (E_{c}+c_4) \,, \\
   p_2 &= \Big\{ 3E_{ddc} k^2
      - 4 \left[ E_c(c_4+2E_c) - 6E_{cc}\right]\left[4E_{ccc}-E_{cd}\right]
      \\ & \quad \quad
      + \left[ 3 (4E_{ccc}-E_{cd})E_d + 2c_4(E_d^2-4E_{ccd}+E_{dd})
        + E_c(2E_d^2-4E_{ccd}+E_{dd})\right] k
      \\ & \quad \quad
      - 8 \left[ E_c^3-3E_{cc}E_c + c_4(E_c^2-2E_{cc})\right] E_d
      \Big\}
      \\ & \quad
      \times \left[ 4c_4 (E_c^2-4E_{cc}-E_dk) \right]^{-1}
      \,, \\
   p_1 &= \Big\{ \left(-2E_{d}^2-4 E_{ccd}+E_{dd}\right)E_{c}^2
         + \left[E_{d} (-4E_{ccc}+E_{cd}-2c_4E_{d})+E_{ddc}k\right] E_{c}
      \\ & \quad \quad
      + 4E_{cc} (E_{d}^2+4E_{ccd}-E_{dd})
         + 2c_4 (-4E_{ccc}E_{d}+E_{cd}E_{d}-E_{ddc}k) \Big\}
      \\ & \quad
      \times \left[ c_4(E_{c}^2-4E_{cc}-E_{d}k)\right]^{-1}
      \,, \\
   p_0 &= \Big\{ 8E_{ddc}E_{c}^2
         + (2E_{d}^3-4E_{ccd}E_{d}+E_{dd}E_{d}-4c_4E_{ddc})E_{c}
      - 24 E_{cc}E_{ddc}
      \\ & \quad \quad
      + E_{d} \left[ 12E_{ccc}E_{d}+2c_4(E_{d}^2+4E_{ccd}-E_{dd})
         - 3(E_{cd}E_{d}+E_{ddc}k) \right] \Big\}
      \\ & \quad
      \times \left[ c_4(-3E_{c}^2+12E_{cc}+3E_{d}k)\right]^{-1}
      \,,
\end{aligned}
\end{equation}
where $c_4$ is either of the two roots of
\bea
c_4^2 &=& k_1^2 c_1^2 + k_2^2 c_2^2 + k_3^2 c_3^2  - k_1 k_2 c_1 c_2 - k_2 k_3 c_2 c_3 - k_3 k_1 c_3 c_1  \,.
\eea

One can readily check that all the solutions presented in the previous sections follow from this
quartic polynomial by imposing the appropriate conditions discussed in section \ref{sec:ansatz}.
In addition, \bref{generalH} leads to interesting generalisations.
Recall that in the class of solutions with $B_6=KE_4\times KE_2$,
we found positive definite metrics in the cases $B_6=KE_4^+\times H^2$,
$B_6=KE_4^-\times S^2$, and $B_6=KE_4^+\times T^2$. For these, \bref{generalH}
leads to a generalisation where the $KE_4$ splits into two $KE_2$ spaces with different radii
(but with both still having the same sign of the curvature). We leave a more detailed analysis of
these generalisations to future work.

On the other hand, the quartic polynomial \bref{generalH} also
provides a generalisation of the $KE_6^{\pm}$ solutions presented in
section \ref{sec:KE6}, where the $KE_6$ splits into three $KE_2$
spaces with different radii but still with the same sign of the
curvature. Recall that we only found singular metrics in the $KE_6^+$
class. Unfortunately, the situation does not seem to improve by
considering the more general solution \bref{generalH} as we have not
been able to find any regular metrics in the $KE_2^+ \times KE_2^+
\times KE_2^+$ class, nor any positive definite metrics in the $KE_2^-
\times KE_2^- \times KE_2^-$ class. However, we have not carried out a
systematic analysis of all possibilities.


\section{Solutions of type IIB String Theory}
\label{iib}


In section~\ref{sec:KE4xT2-class} above we presented
two classes of solutions with $B_6=KE^+_4\times T^2$: the $c_1=0$
solutions with $d_3=0,\pm1$ and the $c_1=1$ solutions parametrised by
$a$. By dimensional reduction on one
leg of $T^2$, and T-duality on the other, these can be transformed
into new solutions of type IIB supergravity with only the five-form
flux excited. As such these should provide new examples of the AdS-CFT
correspondence where the $N=(0,2)$ two-dimensional CFT arises from a
configuration of D3-branes.

The IIB duals of the second family with $c_1=1$
were discussed in some detail in~\cite{GMMW2}. So below we mostly
focus on the $c_1=0$ solutions, analysing the regularity of the solutions,
the conditions for integral flux and deriving an expression for
the central charge of the dual CFT. We show that this family includes
a solution first constructed in~\cite{naka} which describes D3-branes
wrapping a K\"ahler two-cycle in a Calabi-Yau four-fold.
Turning to the $c_1\neq 0$ solutions, we demonstrate how quantisation of the
$G_4$ flux in eleven dimensions is, as expected, related to the
regularity and flux quantisation in the dual type IIB configuration
discussed in~\cite{GMMW2}.

We also note here that the following analysis can also be applied to the
analogous solutions in $B_6=KE_2\times KE_2\times T^2$ class,
but we shall leave the details to future work.


\subsection{$c_1=0$ solutions}
\label{ceq0}


Dualising the solution~\eqref{c1=0} to type IIB, using the formulae at
the end of section~\ref{sec:ansatz}, one finds the metric takes the form
\bea
ds^2_\mathrm{IIB}&=&ds^2(AdS_3)+\frac{3}{4}ds^2(H_2)+\frac{9}{4}ds^2(SE_5) \,,
\label{naka}
\eea
where\footnote{We have rescaled the metric by a factor of
  $(3/8)^{1/2}$ and hence the five-form by $(3/8)^2$ with respect to
  our general IIB formula \bref{general-IIB}. We have also rescaled
  $z$ by a factor of two.}
\be
ds^2(H_2)=\left(r^2-\frac{16d_3}{3}\right)^{-1}dr^2
    +  \left(r^2-\frac{16d_3}{3}\right)dz^2 \,,
\ee
is the constant curvature metric on $H^2$ (irrespective of whether
$d_3=0,\pm 1$), and
\be
\label{SE}
ds^2(SE_5)=\frac{1}{6}\left[ds^2(KE_4^+)+\frac{2}{3}(D\psi+r dz)^2\right] \,.
\ee
Finally, the five-form flux reads, for the case $d_3=0$,
\be
g_s F_5={3\over 32} \left(-J \wedge J +J\wedge dr\wedge dz\right)
   \wedge (D\psi+r dz)
   +\left(\frac{1}{4}J-2dr \wedge dz\right) \wedge \Vol{AdS_3} \,.
\label{naka-5form}
\ee
All other fluxes vanish.

We first observe that, at fixed $z$ and for a given $KE_4^+$ manifold,
$ds^2(SE_5)$ is a regular Sasaki-Einstein metric. We will see shortly
that we can also consider quasi-regular Sasaki-Einstein manifolds
for which $KE^+_4$ is an orbifold. We also note that in order to
obtain a compact solution, we need to take the quotient $H^2/\Gamma$
by an element $\G$ of $SL(2,\bbZ)$ (the Killing spinors are independent of
the $H^2$ coordinates and are therefore preserved in the quotient
procedure).

It is also interesting to point out that that irrespective of whether $d_3=0,\pm 1$
we get the same type
IIB solution. We noted earlier that in the $D=11$ solutions $d_3=0$ was
singular, but $d_3=\pm 1$ were regular. This is not a contradiction,
because in obtaining the $D=11$ solutions from the type IIB solutions
we are T-dualising on different $U(1)$ directions of $H^2$.

We now turn to examine the conditions for the general solutions to be
globally well defined. Note that~\bref{naka} is the metric on a compact
seven-manifold $\CM_7$ which is, locally, a $U(1)$ fibration over
$\mathcal{B}_6=(H^2/\G) \times KE_4^+$. The $U(1)$ fibration is
characterised by the first Chern class $c_1(\CM_7)$. If we let $\psi$
have period $\Delta \psi=2\pi l$, then from~\bref{SE} we find
\be
c_1(\CM_7) = l^{-1}\CR_{KE_4^+} + l^{-1}\Vol{H^2/\G} \,.
\ee
We will have a proper $U(1)$ fibration if $c_1(\CM_7) \in
H^2(\mathcal{B}_6,\bbZ)$. Since the cohomology of $KE^+_4$ and hence
of $\mathcal{B}_6$ contains no torsion classes, this global condition
is equivalent to the periods of $c_1(\CM_7)$ being
integral. Explicitly, a convenient basis for $H_2(B_6,\bbZ)$
is provided by $\{H^2/\G,\Sigma_a\}$, where $\Sigma_a$ is a basis of
$H_2(KE_4^+,\bbZ)$, then we require
\bea
\label{globalcon}
{1\over 2\pi}\int_{\Sigma_a} c_1(\CM_7) \,
   = \, \frac{mn_a}{l}\, \in \bbZ
\sac
{1\over 2\pi}\int_{H^2/\G} c_1(\CM_7) \,
   = \, \frac{\chi}{l}\,  \in \bbZ \,,
\eea
where $\chi$ is the Euler number of the Riemann surface $H^2/\G$, $m$
is the Fano index of the $KE_4^+$ space, and the
integers $n_a$ are coprime (see Appendix~\ref{sec:appendix-KE4} for
further details). Therefore, we deduce that the maximum value
that $l$ can take is
\be
l = \mbox{hcf}\{m,|\chi|\} \,.
\ee
For example, if $KE_4^+=\bbC \bbP^2$, then $m=3$. If, furthermore, the
Euler number of $H^2/\G$ is divisible by 3, then we can take
$l=3$. For fixed $z$, the $S^1$ fibration over the $KE_4^+$ is then,
topologically, an $S^5$. However, for a general choice of $H^2/\G$,
the largest possible $l$ is $l=1$, which leads to $S^5/\bbZ_3$.

By considering, for example, the family of $Y_{p,q}$ Sasaki-Einstein
metrics~\cite{GMSW2} on $S^3\times S^2$, one can follow a similar
argument to show that $\mathcal{M}_7$ is still regular when $KE_4^+$
is replaced by the orbifold base of a quasi-regular
$Y_{p,q}$. However, the construction does not appear to work for
irregular $Y_{p,q}$ metrics. To check the regularity it is natural to
change coordinates on $Y_{p,q}$ to the parameterisation where
$Y_{p,q}$ is manifestly a $U(1)$ bundle over a base which is itself an
$S^2$ bundle over $S^2$. Given the form of~\eqref{naka}
and~\eqref{SE}, in this coordinate system, $\mathcal{M}_7$ similarly
becomes a $U(1)$ fibration over an $S^2$ fibration now over $S^2\times
H^2/\Gamma$. One can then check the regularity of the $U(1)$ and $S^2$
fibres over $H^2/\Gamma$. One finds that the whole space can be made
regular, but generically one must make the size of the $U(1)$
fibration non-maximal, that is consider $Y_{np,nq}$ for some positive
integer $n$ and $(p,q)$ coprime (that is a $\bbZ_n$ quotient of
$Y_{p,q}$). One would expect that a similar construction would work
for any quasi-regular Sasaki-Einstein space.

Let us now compute the central charge for these solutions. We will
follow the same steps as in~\cite{GMMW2}.
We first reinstate dimensions by rescaling the metric and background,
\bea
d\tilde{s}^2=L^2 ds^2 \sac \tilde{F}_5=L^4 F_5 \,.
\eea
The quantisation condition for the five-form in type IIB is
\be
\label{defi}
N(D) = {1 \over (2\pi l_s)^4} \int_{D} \tilde{F}_5 \,\, \in \,\, \bbZ\,,
\ee
where $D$ is any five-cycle of $\CM_7$.

Now $H_5(\CM_7,\bbZ)$ is generated by two types of five-cycle:
$D_0$, the $U(1)$ fibration over the $KE_4^+$ space, and
$D_a$, the five-cycles obtained from the $U(1)$ fibration over $H^2/\G
\times \Sigma_a$. From \bref{defi}, we obtain
\be
\label{flux-units}
N(D_0) = -{3 L^4 \over 64 \pi l_s^4 g_s } \, M l \sac
N(D_a) =  {3 L^4 \over 64 \pi l_s^4 g_s } \, \chi \,l  \, m n_a \,,
\ee
where
\bea
M = {1 \over (2\pi)^2} \int_{KE_4^+} \CR \wedge \CR \,.
\eea
Noting that $M$ is always divisible by $m^2$, and hence in particular by $m$,
the condition that all the fluxes are the minimal possible integers becomes
the following quantisation condition on the $AdS_3$ radius:
\bea
{3 L^4 \over 64 \pi g_s l_s^4} = {n \over m l h} \sac
h=\mbox{hcf}\{\tfrac{M}{m},|\chi|\} \,,
\eea
leaving $N(D_0)=-\tfrac{M}{m h}n$ and $N(D_a)=\tfrac{\chi}{h} n_a n$.
For $n=1$ we obtain the minimal D3-brane setup that creates the background, with
higher values of $n$ corresponding to $n$ copies of this minimal system.

We can now determine the central charge $c$ of the dual two-dimensional SCFT.
It is well known~\cite{Brown:1986nw} that $c$ is fixed by the $AdS_3$
radius $L$ and the Newton constant $G_{(3)}$ of the effective
three-dimensional theory obtained by compactifying type IIB
supergravity on $\CM_7$:
\be
c = {3 L \over 2 G_{(3)}} \,.
\ee
Using the same conventions as in~\cite{GMMW2}, we obtain the following
expression:
\bea
c = {36M |\chi|\over m^2 h^2 l } n^2 \,,
\eea
which, remarkably, gives an integer number irrespective of the choice
of $KE_4^+$ and $H^2/\G$.

In the special case that we choose $KE_4^+=\bbC\bbP^2$ we find that we
have recovered the type IIB solution
that corresponds to D3-branes wrapping a holomorphic $H_2$-cycle (or
$H^2/\Gamma$-cycle) inside a Calabi-Yau four-fold. This solution was
first found in~\cite{naka} and generalises the solutions of~\cite{mn}.
From this perspective we have shown that the solutions of~\cite{naka}
(at least for non-compact $H^2$) can be dualised and uplifted to
regular solutions in $D=11$. Note that the $S^5$ in this solution is
the $S^5$ that surrounds the D3-branes. We have thus also shown
that we can replace\footnote{We can analogously replace the $S^7$ in
  the $D=11$ solutions describing membranes wrapping holomorphic
  curves in Calabi-Yau five-folds that were constructed
  in~\cite{Gauntlett:2001qs}, by a seven-dimensional regular or
  quasi-regular Sasaki-Einstein manifold.}
this $S^5$ with any regular or quasi-regular Sasaki-Einstein metric.


\subsection{$c_1\ne 0$ solutions}
\label{cneq0}


The reduction to type IIB of the $B_6=KE^+_4\times T^2$ solutions with
$c_1\neq 0$ gives precisely the solutions we presented in
\cite{GMMW2}. The metric~\eqref{IIBmetric} on the seven-dimensional
IIB manifold $\CM_7$ is given by
\begin{equation}
\label{cneq0metric}
   ds^2({\cal M}_7) = {3 \over 8y} ds^2_{KE_4}
      + {9 dy^2\over 4q(y)} + {q(y) D\psi^2\over 16 y^2(y^2-2y+a)}
      + {y^2-2y+a \over 4 y^2} Dz^2 \,,
\end{equation}
where, as above, $q(y)=4 y^3 -9 y^2 +6 a y-a^2$ and
$Dz=dz+A_1$, with $F_2=dA_1$ given in~\eqref{c1neq0} or explicitly
\begin{equation}
   Dz = dz - \frac{a-y}{2(y^2-2y+a)}D\psi \,.
\end{equation}
The global analysis of~\cite{GMMW2} proceeded in two steps. First we
showed that $(y,\psi)$ form an $S^2$ fibration $\mathcal{B}_6$ over
the base $KE_4$.  We then showed that $z$ formed a $U(1)$ fibration
over $\mathcal{B}_6$.

In this section we will show how this analysis translates into
conditions on the M-theory solution. Note first that the $S^2$ of the
M-theory solution is only fibred over the $KE^+_4$ part of
$B_6=KE^+_4\times T^2$. From this perspective we can write
$\CM_8=\mathcal{B}_6\times T^2$ where $\mathcal{B}_6$ is the same
manifold that appears in the IIB solution: $(y,\psi)$ form an $S^2$
fibration over $KE_4$ in eleven dimensions. Thus the global analysis
of $\mathcal{B}_6$ is the same. Explicitly, it is regular and compact
for values of $a\in(0,1)$ if the range of $y$ is restricted to lie
between the first two roots of $q(y)$.

The second part of the IIB global analysis performed in~\cite{GMMW2}
translates, however, not into geometry but quantisation of the $G_4$
flux in eleven dimensions, as we will now show. To this aim, we first
reinstate dimensions by rescaling the eleven-dimensional metric and
background,
\bea
d\tilde{s}^2_{11}& =& L^2 ds^2_{11} \sac \tilde{G}_4=L^3 \, G_4 \,.
\eea
Note that this implies an analogous rescaling the IIB metric and five form.
As discussed in section~\ref{flux}, for all our examples, the
quantisation condition for $\tilde{G}_4$ is that for any four-cycle
$\mathcal{D}$ we have
\bea
\label{11d-flux}
N(\mathcal{D})
   = {1 \over (2\pi l_P)^3}\int_{\mathcal{D}} \tilde{G}_4
   = {L^3 \over (2\pi l_P)^3}\int_{\mathcal{D}} G_4
   \,\, \in \,\, \bbZ \,.
\eea
Recall that, upon reduction and T-duality along the $T^2$, the
surviving direction of the torus becomes the $z$-circle fibred over
$\mathcal{B}_6$. In~\cite{GMMW2} we denoted by $l$ the resulting
radius of this circle measured in units of $L$, and we studied the
conditions for the fibration to be a proper $U(1)$ bundle: namely,
that the periods of its first Chern class be integer numbers. Thus, in
the IIB language, we require
\bea
P(C) = {1 \over 2\pi l} \int_{C} F_2 \,\, \in \,\, \bbZ
\label{examp}
\eea
where $C$ is an two-cycle in $H_2(\mathcal{B}_6,\bbZ)$.

We can now use the standard relations between eleven-dimensional and
type IIB parameters (see appendix~\ref{sec:appendix-orientations})
\begin{equation}
\label{relations}
   R_\mathrm{IIB} = lL = l_P^3/R_1R_2 \,, \qquad
   l_s^2 = l_P^3/R_1 \,, \qquad
   g_s = R_1/R_2 \,.
\end{equation}
From the first relation we have
\begin{equation}
\begin{aligned}
   P(C) &= \frac{LR_1R_2}{2\pi l_P^3} \int_{C} F_2
      = \frac{L^3\,\mathrm{Vol}(T^2)}{(2\pi l_P)^3} \int_{C} F_2
      = {L^3 \over (2\pi l_P)^3} \int_{C\times T^2} G_4 \\
      &= N(C \times T^2) \,.
\end{aligned}
\end{equation}
where we have used $\mathrm{Vol}(T^2)=(2\pi)^2R_1R_2/L^2$ since the
volume is measured in units of $L$.
Thus the integrality of the IIB periods is equivalent to the quantisation
of the flux through the four-cycles
$\mathcal{D}=C\times T^2$ in the M-theory solution.

In the IIB solution we also needed to ensure that the flux of $F_5$
through any 5-cycle $D$ of $\CM_7$ is appropriately quantised. The
only non-trivial cycles arise as $S^1$ fibrations over
a non-trivial four-cycle $\mathcal{D}$ in
$H_4(\mathcal{B}_6,\bbZ)$. Now, recall that the IIB 5-form is
\be
\label{F5G4}
g_s F_5 = L^4 \, (1+\star)  \,\, \left[ F_4 \wedge Dz \right]\,,
\ee
where $z$ parametrises the $S^1$. The quantisation condition reads
\bea
N^{\mathrm{IIB}}(D) =
   \left({L \over 2\pi l_s }\right)^4
      \int_{D} g_s^{-1} F_4 \wedge dz
   = {L^4 l \over (2\pi)^3 g_s l_s^4 } \int_{\mathcal{D}} G_4 \, \,,
\eea
for $\mathcal{D}\in H_4(\mathcal{B}_6,\bbZ)$ and where in going to the
final expression we have integrated over the $S^1$ fibre and used the
expression~\eqref{F5G4}. Using the relations~\eqref{relations} we
directly see that
\be
N^{\mathrm{IIB}}(D) = N(\mathcal{D}) \,,
\ee
so that the quantisation of $F_5$ is equivalent to the quantisation of
$G_4$ through four-cycles $\mathcal{D}$ in $\mathcal{B}_6$.
Notice that the ratio of the two radii of the torus is unfixed,
corresponding to the fact that the IIB dilaton can take any constant
value.


\section{Non-Compact Solutions}
\label{sec:noncompact}


So far we have focussed on solutions where the internal space is
compact as this leads to new examples of the AdS-CFT
correspondence. However, our analysis can also be used to find new
solutions where $\CM_7$ is non-compact. In this section we will
initiate a study of such solutions, restricting our attention to the
class of type IIB solutions with $c_1\ne 0$ that were first presented
in~\cite{GMMW2} and briefly discussed above in section~\ref{cneq0}.

It is convenient to introduce the coordinates $\psi=\psi'-z'$, $z=2z'$
so the local class of solutions~\eqref{cneq0metric} parametrised by
$a$ can be written
\bea
ds^2 &=& \frac{9}{4}L^2 \left[ \frac{4y}{9}ds^2({AdS_3}) +
  \frac{y}{q(y)} dy^2+\frac{q(y)}{9 y^2} dz'^2
  + {1 \over 6} ds^2_{KE_4} + {2 \over 3} \left(D\psi' - A \right)^2
\right] \,,
\nn
A &=& {a \over y} dz' \,.
\eea
with
\begin{equation}
\begin{aligned}
   g_s L^{-4} F_5 &= J  \wedge\left[
      {3 \over 32} J \wedge (D\psi'-A)
      + {3 a \over 34 y^2} dy \wedge D\psi' \wedge dz' \right]
      \\ & \qquad \qquad
      {} + \Vol{AdS_3}\wedge \left[
         2y \,  dy \wedge dz' - \frac{a}{4} \, J \right] \,.
\end{aligned}
\end{equation}
For the compact solutions, $y$ ranged between $y_1$ and $y_2$, the two
smallest roots of the cubic $q(y)$. For the non-compact solutions, we
instead take $y_3\le y<\infty$, where $y_3$ is the largest root of the
cubic.

Let us first consider the case when $a=0$ giving $y_1=y_2=0$ and
$y_3=9/4$. By implementing the coordinate change $y=(9/4)\cosh^2\rho$,
the metric becomes
\bea
ds^2 &=& \frac{9}{4}L^2 \left[ \cosh^2\rho ds^2({AdS_3}) + d\rho^2
   +\sinh^2\rho dz' + {1 \over 6} (ds^2_{KE_4} + {2 \over 3}
   \left(D\psi'\right)^2) \right] \,,
\eea
Remarkably, we have just recovered the $AdS_5\times SE_5$ solutions of
type IIB supergravity, where $SE_5$ denotes a five-dimensional
Sasaki-Einstein manifold. In particular, if $KE_4=CP^2$  we obtain
$AdS_5\times S^5$.

We next observe that for general $a$, as $y\to\infty$ the solution
behaves as if $a=0$ and hence the solutions are all asymptotic to
$AdS_5\times SE_5$. Furthermore, for generic $a$ (not equal to 0 or 1)
when the three roots $y_i$ are distinct, we see that as $y$ approaches
$y_3$ the potential conical singularity can be removed by taking the
period of $z'$ to be $2\pi[6y_3^{3/2}/q'(y_3)]$. With this period the
non-compact solutions are regular: they are fibrations of $SE_5$ over a
five-dimensional space which is a warped product of $AdS_3$ with a
disc parametrised by $y,z'$.

To interpret these solutions we consider for simplicity the case when
$SE_5=S^5$. There are probe D3-branes in $AdS_5\times S^5$ whose
world-volume is $AdS_3\times S^1$. Following~\cite{kr}, in terms of
intersecting branes, one such configuration~\cite{st} is just two flat
D3-branes intersecting over a string, where the geometry $AdS_5\times
S^5$ corresponds to the near-horizon limit of one of the branes, while the
second brane is treated as a probe. However, such a configuration
preserves $\frac{1}{4}$ of the supersymmetry of Minkowski space,
whereas our configurations preserve $\frac{1}{16}$th. A configuration
with the correct supersymmetry would be four D3-branes intersecting
over a string, with the geometry corresponding again to the
near-horizon limit of one of the branes. Specifically, if
$(z_1,z_2,z_3)$ are complex coordinates in the $\bbR^6$ space
transverse to this background brane, the other three probe branes could
lie in the orthogonal holomorphic two-planes $z_1=z_2=0$, $z_2=z_3=0$
and $z_3=z_1=0$. These probe branes are a generalisation of those
studied in~\cite{kr} which corresponded to defect CFTs. It is natural
to
interpret our new solutions as the back-reacted geometry of
such probe branes and dual to a four-dimensional $N=4$
super Yang-Mills theory coupled to string-like defects which preserve
the $N=(0,2)$ two-dimensional superconformal subgroup of $PSU(2,2|4)$.
One might expect the back-reacted geometry of such branes
to be localised in $CP^2$ corresponding to the positions of the three
probe branes. However, in our solutions the $CP^2$ is still
manifest. Hence our geometries seem to correspond to probe D3-branes
that have been ``smeared'' over the $CP^2$. In terms of intersecting
branes, instead of three probe branes, one is considering a uniform
superposition of flat probe branes spanning all holomorphic
two-planes\footnote{A given plane is parameterised by
  $z_i=\lambda_iw$ for generic constants $\lambda_i$ and parameter
  $w\in\bbC$.}  in $\bbC^3$.  More generally, the solutions where the
$S^5$ factor is replaced by $SE_5$ can similarly be interpreted as the
gravity duals of a general $N=1$ SCFT coupled to string-like defects.

We make a final observation about the $a=1$ case, for which $q(y)$ has
a double root at $y=1$. By expanding the solution near $y=1$ we find,
again remarkably, that the solution is asymptotically approaching the
solutions discussed in section~\ref{ceq0}. In particular, for the special
case when $KE_4=CP^2$, this is the solution found in~\cite{naka} that
describes the near horizon limit of D3-branes wrapping a holomorphic
$H^2/\Gamma$ in a Calabi-Yau four-fold. Thus, in this special case,
our full non-compact solution interpolates between $AdS_5\times S^5$
and the solution~\cite{naka}, while preserving an $AdS_3$ factor.


\section{Summary and Conclusions}
\label{sec:conclusions}


In this work we have found new infinite classes of supersymmetric
warped $AdS_3\times \CM_8$ solutions of eleven-dimensional
supergravity. The new compact solutions are all $S^2$ bundles over
six-dimensional base spaces $B_6$ which are products of
K\"ahler-Einstein spaces. The explicit solutions are obtained by
solving the single second-order differential equation \bref{Heq}, for
which we have found the most general {\it polynomial} solution. The
most general polynomial solution arises for the case when
$B_6=KE_2\times KE_2\times KE_2$ and gives the quartic solution
\bref{generalH}. The solutions for $B_6=KE_4\times KE_2$ and
$B_6=KE_6$ can then be obtained from this general solution as special
cases. The new compact regular classes of solutions can be summarised
as follows:
\begin{itemize}
\item
The first class is when $B_6=KE_2^+\times KE_2^+\times H^2$. In the
special limit where the two $KE_2^+$ radii coincide, the same class
can also describe solutions with $B_6=KE_4^+\times H^2$, as discussed
in section~\ref{sec:inverse-GKW-class}.
\item
The second class is when $B_6=KE_2^- \times KE_2^- \times S^2$. In the
special limit where the two $KE_2^-$ radii coincide, this class can
also describe solutions with $B_6=KE_4^-\times S^2$, as discussed in
section \ref{sec:GKW-class}. The latter include the
solution~\cite{GKW} originally found in gauged supergravity,
describing M5 branes wrapping a K\"ahler four-cycle in a Calabi-Yau
four-fold.
\item
The third class is when $B_6=KE_2^+ \times KE_2^+ \times T^2$. In the
special limit where the two $KE_2^+$ radii coincide, this class can
also describe solutions with $B_6=KE_4^+\times T^2$, as discussed in
section \ref{sec:KE4xT2-class}. This class is particularly interesting
because it leads to type IIB backgrounds with constant dilaton and
only 5-form flux. Focussing on the IIB solutions arising from the
$B_6=KE_4^+\times T^2$ case, in section \ref{sec:Naka-class}, we
showed how these lead to generalisations of the solutions
corresponding to D3 branes wrapping an $H^2/\Gamma$ in a $CY_3$ found
in~\cite{naka}, whereas in section \ref{sec:IIB-class} we showed how
to recover the infinite IIB families presented in~\cite{GMMW2}. The
IIB solutions solutions arising from the $B_6=KE_2^+\times KE_2^+
\times T^2$
case provide generalisations that would be interesting to
explore further.
\end{itemize}

The general polynomial solution to the differential equation
\bref{Heq} thus gives rise to an extraordinarily rich five-parameter
family of solutions. The most general solution involves eight
integration constants and it would be interesting to know if this
larger family includes any additional regular solutions.

Despite the long expressions involved in the most general polynomial
solution, we gave a simple argument in section~\ref{sec:no-conical}
(covering most cases) that the metric is regular. Therefore, the solutions
presented here always lead to good eleven-dimensional supergravity
backgrounds. However, as M-theory backgrounds, we still need to make
sure that the fluxes of the four-form field strength are integral. We
have only discussed the implications of this condition for the
solutions where $B_6=KE_4^+\times T^2$. As we discussed in section
\ref{sec:IIB-class} this requires appropriately discretising both the
volume of the torus and the parameter of the solution, leading to an
infinite discrete series of $AdS_3/CFT_2$ examples. Though the most
general case is more difficult to analyse, we expect that most, if not
all, of the parameters of the solution will also need to be
discretised. It would be interesting to check this expectation since
if it were not true, we would be led to predict the existence of
exactly marginal deformations of the dual CFTs.

It would be very interesting if the dual conformal field theories to
our new solutions could be identified. The solutions in the third class
above that have a type IIB description seem the most promising, since
the CFTs must arise from the gauge theories living on D3-branes. A key
check of any proposal will be to recover the central charges that were
calculated here and in~\cite{GMMW2} for the case when $B_6=KE_4\times
T^2$.

We also found some intriguing non-compact solutions. In particular,
there were type IIB solutions in the $B_6=KE_4\times T^2$ class
that, in the simplest case, appear to be dual to the two-dimensional
defect CFTs arising on probe D3-branes with world-volume $AdS_3\times
S^1$ embedded in $AdS_5\times S^5$. More generally, the $S^5$ factor
could be replaced with $SE_5$ such that the geometries appear dual to
defect CFTs in more general $N=1$ field theories. This also seems to
be a profitable avenue for further investigation.


\subsection*{Acknowledgements}
We would like to thank Dmitriy Belov, Nakwoo Kim, Kostas Skenderis,
James Sparks and Marika Taylor for discussions. OC is supported by
EPSRC. DW is supported by the Royal Society through a University
Research Fellowship.


\appendix


\section{Conditions for Supersymmetry}
\label{sec:appendix}


In this appendix we determine the conditions for the ansatz
\bref{an1}--\bref{an3} to be a supersymmetric solution to the
equations of motion of $D=11$ supergravity. For completeness and by
way of comparison, we shall consider two equivalent approaches. First
we substitute directly into the Killing spinor equations, given a
particular ansatz for the Killing spinor. This is the most
straightforward approach and gives the explicit dependence of the
spinor on the coordinates. In the second approach we use G-structure
techniques. In particular, we follow the general analysis
of~\cite{GMMW1}. This has the advantage of giving generic conditions
for a general class of $AdS_3$ compactifications. Substituting our
particular ansatz then gives a comparatively easy way of obtaining the
relevant differential equations. It also motivates a particular gauge
choice for the function $f_3$.

We shall use the conventions of~\cite{Gauntlett:2002fz}. In particular
the Killing spinor equations are
\bea
\left(\nabla_b +{1 \over 288} \left[ \G_{b}~^{a_1...a_4}-8 \delta_{b}^{a_1} \G^{a_2...a_4} \right] G_{a_1...a_4} \right)\e =0 \,.
\eea
In addition we need to ensure that the Bianchi identity and the
equations of motion for the four-form $G_4$ are satisfied
\be
dG_4=0\sac d*G_4+\frac{1}{2}G_4\wedge G_4=0 \,.
\ee
As we will see below, the Killing spinors of the
$AdS_3$ solutions we construct define a preferred local $SU(3)$
structure. Consequently, by the arguments of \cite{oap}, it is
sufficient to impose just the Bianchi identity, since all field
equations are then identically implied by the
supersymmetry conditions.


\subsection{Killing spinor analysis}
\label{app:KS}

Let us start with the Bianchi identity. Given our ansatz, it implies the
following three equations
\bea \label{basic-bianchis}
g_1'=k_3 g_5+k_2 g_6 \sac g_2'=k_3 g_4+k_1 g_6 \sac g_3' =k_2 g_4+k_1 g_5  \,.
\eea
The Killing spinor equation is more involved.
We will use the following orthonormal
frame
\bea
e^\a &=& \w \te^\a \sac
e^a = A_1 \te^a \sac
e^i = A_2 \te^i \sac
e^m = A_3 \te^m  \,,\nn
e^r &=& B dr \sac
e^{\psi} = C (d\psi+\tP) \nonumber \,,
\eea
with
\be
A_i=\w h_i^{1/2} \sac B=\w f_3^{1/2}\sac  C=\w f_4^{1/2} \,,
\ee
where we have introduced the index notation
\bea
AdS_3 :\, \a=\{\tilde{0},\tilde{1},\tilde{2}\} &\sac&
C_1  :\, a = \{1,2\}\,, \nn C_2  :\, i = \{3,4\} &\sac& C_3  :\, m=\{5,6\} \,.
\nonumber
\eea
The K\"ahler forms for the $C_i$ are given by
\be
J_1=e^1\wedge e^2 \sac J_2=e^3\wedge e^4 \sac J_3=e^5\wedge e^6 \,.
\ee
Our orientation is fixed by $\epsilon_{\tilde{0}\tilde{1}\tilde{2}123456r\psi}=1$, where
the indices refer to the vielbein basis.
It will be useful to
define
\be
\gamma_9\equiv\Gamma_{123456r\psi} =\Gamma_{\tilde{0}\tilde{1}\tilde{2}} \,.
\ee

A straightforward calculation then shows that the Killing spinor equations take the following form, where
all indices are tangent space indices:
\begin{small}
\bea
0&=& \bigg\{{1 \over \w} \hat{\nabla}_\a +{\w' \over 2B\w} \G_{\a r} +
{1 \over 12} \G_{\a} \left[ {g_3\over 4 A_1^2 A_2^2} J_1 \G J_2 \G +
{g_1\over 4 A_2^2 A_3^2} J_2 \G J_3 \G +{g_2\over 4 A_3^2 A_1^2} J_3 \G J_1 \G \right.
\nn
&& \left. + {1 \over BC} \left({g_4 \over 2 A_1^2} J_1 \G+{g_5 \over 2 A_2^2} J_2 \G+{g_6 \over 2 A_3^2} J_3 \G\right) \G_{r \psi}
\right] +{g_7 \over 6 B \w^3} \G_{\a r} \g_9\bigg\}\epsilon \,,
\nn
0&=& \bigg\{{1 \over A_1} \left( \hat{\nabla}_a - P_a \pa_{\psi} \right) + {A_1' \over 2 B A_1} \G_{a r} -
{k_1 C \over  4 A_1^2} J_{ab} \G_{\psi b} \nn
&&+{1 \over 12} \G_a \left[ {g_1 \over 4 A_2^2A_3^2} J_2 \G J_3 \G
 +  {1 \over BC} \left( {g_5 \over 2 A_2^2} J_2\G + {g_6 \over 2 A_3^2} J_3\G\right) \G_{r \psi}
-{g_7 \over B \w^3} \G_r \g_9 \right]
\nn
&&  - {1 \over 6} J_{ab}\G_b \left[ {g_3 \over 2 A_1^2 A_2^2} J_2 \G
+{g_2 \over 2 A_3^2 A_1^2} J_3 \G + {g_4\over BC A_1^2} \G_{r\psi} \right]\bigg\}\e \,,
\nn
0&=& \bigg\{{1 \over A_2} \left( \hat{\nabla}_i - P_i \pa_{\psi} \right) + {A_2' \over 2 B A_2} \G_{i r} -
{k_2 C \over  4 A_2^2} J_{ij} \G_{\psi j} \nn
&&+{1 \over 12} \G_i \left[ {g_2 \over 4 A_3^2A_1^2} J_3 \G J_1 \G
 +  {1 \over BC} \left( {g_4 \over 2 A_1^2} J_1\G + {g_6 \over 2 A_3^2} J_3\G\right) \G_{r \psi}
-{g_7 \over B \w^3} \G_r \g_9 \right]
\nn
&&  - {1 \over 6} J_{ij}\G_j \left[ {g_3 \over 2 A_1^2 A_2^2} J_1 \G
+{g_1 \over 2 A_2^2 A_3^2} J_3 \G + {g_5\over BC  A_2^2} \G_{r\psi} \right]\bigg\}\e \,,
\nn
0&=& \bigg\{{1 \over A_3} \left( \hat{\nabla}_m - P_m \pa_{\psi} \right) + {A_3' \over 2 B A_3} \G_{m r} -
{k_3 C \over 4 A_3^2} J_{mn} \G_{\psi n} \nn
&&+{1 \over 12} \G_m \left[ {g_3 \over 4 A_1^2A_2^2} J_1 \G J_2 \G
 +  {1 \over BC} \left( {g_4 \over 2 A_1^2} J_1\G + {g_5 \over 2 A_2^2} J_2\G\right) \G_{r \psi}
-{g_7 \over B \w^3} \G_r \g_9 \right]
\nn
&&  - {1 \over 6} J_{mn}\G_n \left[ {g_1 \over 2 A_2^2 A_3^2} J_2 \G
+{g_2 \over 2 A_3^2 A_1^2} J_1 \G + {g_6\over BC  A_3^2} \G_{r\psi} \right]\bigg\}\e \,,
\nn
0&=& \bigg\{{1 \over B} \pa_r +{1 \over 12} \G_r \left[{g_3\over 4 A_1^2 A_2^2} J_1 \G J_2 \G +
{g_1\over 4 A_2^2 A_3^2} J_2 \G J_3 \G +{g_2\over 4 A_3^2 A_1^2} J_3 \G J_1 \G \right]
\nn
&&-{1 \over 6 BC} \left[{g_4 \over 2 A_1^2} J_1 \G+{g_5 \over 2 A_2^2} J_2 \G+{g_6 \over 2 A_3^2} J_3 \G\right] \G_{\psi}
+{g_7 \over 6 B \w^3} \g_9\bigg\}\e \,,
\nn
0&=& \bigg\{{1 \over C} \pa_{\psi} +{C' \over 2BC} \G_{\psi r} -\sum_i{k_i C \over 8 A_i^2} J_i \G
\nn
&& +{1 \over 12} \G_{\psi} \left[{g_3\over 4 A_1^2 A_2^2} J_1 \G J_2 \G +
{g_1\over 4 A_2^2 A_3^2} J_2 \G J_3 \G +{g_2\over 4 A_3^2 A_1^2} J_3 \G J_1 \G
- {g_7 \over B \w^3} \G_r \g_9 \right]
\nn
&&+{1 \over 6 BC} \left[{g_4 \over 2 A_1^2} J_1 \G+{g_5 \over 2 A_2^2} J_2 \G+{g_6 \over 2 A_3^2} J_3 \G\right] \G_{r}\bigg\}\e \,.
\label{killing-spinor}
\eea
\end{small}
The hats on the covariant derivatives indicate the covariant derivatives on the $AdS_3$ and $KE_2$ spaces.

To proceed, we assume that the Killing spinors are of the form
\bea
\e = \alpha(r) e^{\b(r) \G_r \g_9} e^{\undos \psi \G_{r\psi} \g_9} \epsilon_0 \,,
\eea
where $\e_{0}$ satisfies the projections
\bea\label{projs}
\G_{12}\e_0=\G_{34}\e_0=\G_{56}\e_0\sac \gamma_9\e_0=\e_0 \,.
\eea
In addition $\e_0$ must satisfy
\bea
\hat{\nabla}_\a\e_0 &=& \undos \G_\a \g_9\e_0
\,,\nn
\hat{\nabla}_a \e_0 &=& \undos P_a \G_{r\psi} \e_0
\sac \hat{\nabla}_i \e_0= \undos P_i \G_{r\psi} \e_0
\sac \hat{\nabla}_m \e_0 = \undos P_m \G_{r\psi} \e_0 \,.
\label{cycle-spinors}
\eea
The first of these equations can be solved using the Killing spinors on $AdS_3$, while
the other three are solved using the Killing spinors on $C_i$.
Note that the integrability conditions for the last three equations are consistent with the projections imposed in \bref{projs}. The projections
imply that we preserve 1/8 of the supersymmetry.
Two of the
four supersymmetries are Poincar\'e supersymmetries and the other two are special conformal supersymmetries.
Writing the metric on $AdS_3$ in horospherical coordinates, the former are eigenstates of $\G_{\rho}$, the $\G$-matrix
along the $AdS$ radial direction~\cite{Lu:1998nu}. But given that all the spinors preserved in our backgrounds are also eigenstates of
$\g_9$, we deduce that those that become Poincar\'e supersymmetries
all have the same chirality with respect to the Minkowski conformal boundary,
and hence the solutions are dual to conformal field theories with $(0,2)$ supersymmetry.

Another interesting property that follows from \bref{cycle-spinors}, and that we have used
repeatedly in the main text, is that {\it the Killing spinors are independent of the coordinates
on the two-cycles $C_i$}. Essentially, the terms on the right hand sides of
\bref{cycle-spinors}, which arise from the connection on the normal bundle to the cycles, cancel the
spin connection terms inside the covariant derivatives $\hat{\nabla}$. Note that this statement
ceases to be true, in general, when any two of the cycles are replaced with a $KE_4$ space.

Plugging our spinor ansatz into the Killing spinor equations \bref{killing-spinor} leads to
two differential equations for $\a(r),\b(r)$ plus a set of algebraic constraints on the spinor.
The equation for $\a(r)$ can be solved exactly,
\be
\alpha(r)=\w^{1/2}(r) \,,
\ee
whereas the equation for $\beta$ reads
\bea
{1 \over B}\b' &=& {1 \over 2\w} - {1 \over 4BC} \left( {g_4 \over A_1^2} + {g_5 \over A_2^2} + {g_6 \over A_3^2} \right) \,.
\label{eq-beta}
\eea
The remaining algebraic conditions (coming from the $AdS_3$, $C_i$ and $\psi$ directions)
can all be written in the form
\begin{small}
\bea
\calp_{AdS_3} \e = \calp_{C_i} \e = \calp_{\psi} \e = 0 \,,
\eea
with
\bea
\calp_{AdS_3} &=&\frac{\w'}{2B\w} +{g_7 \over 6 B \w^3} \g_9 - {1\over 12}\left[{g_3 \over A_1^2 A_2^2} + \mbox{perm} \right] \G_r
\nn&&+\left[{1\over 2 \w} - {1 \over 12 BC} \left( {g_4 \over A_1^2} + \mbox{perm} \right) \right] \G_r \g_9  \,,
\nn
\calp_\psi &=& \left(\frac{1}{2C}-{C k_i \over 4 A_i^2}+{g_7 \over 12 B \w^3} \right)-{C' \over 2BC} \g_9
+{1\over 6 BC}\left( {g_4 \over A_1^2} +\mbox{perm}\right) \G_r \nn
&&-{1 \over 12} \left({g_3 \over A_1^2 A_2^2} + \mbox{perm}\right) \G_r \g_9
\,,
\nn
\calp_{C_1}&=&{1 \over 6} \left[ {-g_1 \over A_2^2 A_3^2} + {2 g_3 \over A_1^2 A_2^2} +{2 g_2 \over A_3^2 A_1^2} \right]
-{1 \over 6BC} \left[ {g_5 \over A_2^2} + {g_6 \over A_3^2} - {2 g_4 \over A_1^2} \right] \g_9
+ {A_1' \over B A_1} \G_r \nn &&- \left[{g_7 \over 6 B \w^3}  +{k_1 C \over 2A_1^2} \right] \G_r \g_9, \nn
\calp_{C_2}&=&{1 \over 6} \left[ {-g_2 \over A_3^2 A_1^2} + {2 g_3 \over A_1^2 A_2^2} +{2 g_1 \over A_2^2 A_3^2} \right]
-{1 \over 6BC} \left[ {g_4 \over A_1^2} + {g_6 \over A_3^2} - {2 g_5 \over A_2^2} \right] \g_9
+ {A_2' \over B A_2} \G_r \nn && - \left[{g_7 \over 6 B \w^3}+{k_2 C \over 2A_2^2} \right] \G_r \g_9,\nn
\calp_{C_3}&=&{1 \over 6} \left[ {-g_3 \over A_1^2 A_2^2} + {2 g_1 \over A_2^2 A_3^2} +{2 g_2 \over A_3^2 A_1^2} \right]
-{1 \over 6BC} \left[ {g_4 \over A_1^2} + {g_5 \over A_2^2} - {2 g_6 \over A_3^2} \right] \g_9
+ {A_3' \over B A_3} \G_r \nn && - \left[{g_7 \over 6 B \w^3}+{k_3 C \over 2A_3^2} \right] \G_r \g_9.\nn\label{eqs}
\eea
\end{small}
Note that the algebraic equations (\ref{eqs}) have all the same structure
\bea
\left( a_0 + a_1 \g_9 + a_2 \G_r +a_3 \G_r \g_9 \right) e^{\b \G_r\g_9} e^{\undos \psi \G_{r\psi}} \e_0 =0 \,.
\eea
After multiplying from the left by $e^{-\b \G_r\g_9}$,
we find that this is solved provided that
\bea
 a_0 +a_1 \cos2\b+a_2 \sin2\b=0  \sac  a_3 +a_2 \cos2\b-a_1 \sin2\b =0 \,.
\eea
In this way, we obtain ten equations from (\ref{eqs}).
To solve the Killing spinor equation we also need to solve an eleventh equation (\ref{eq-beta}).
In addition to solving these eleven differential equations for the metric and four-form functions
we also have a further three differential equations coming from the Bianchi
identities (\ref{basic-bianchis}).

At this stage, the problem still seems formidable. However, we may progress as follows.
To begin we use seven of the eleven equations
to determine the flux functions $g_1,\dots,g_7$ in terms of the metric
functions and their first derivatives. Specifically, we use the conditions arising from $\calp_{C_i}$ and $\calp_\psi$.
In making further progress we found it extremely useful to work in the gauge
\bea \label{gauge}
f_3^{1/2}=\frac{1}{\omega^3\sin2\beta}.
\eea
In this gauge the expressions for the $g_i$ are given by
\begin{small}
\bea
g_1&=& \frac{\w^3}{3 f_3^{1/2} h_1} [  \cot 2\b ( (f_3f_4)^{1/2} (-2 k_1 h_2 h_3+k_2 h_3 h_1+k_3 h_1 h_2)
+ \cos 2\b [ 3 h_1' h_2 h_3\nn&&- (h_1 h_2 h_3)' ] )
+\sin 2 \b (3 h_1' h_2 h_3 -2 (h_1 h_2 h_3)' -6 h_1h_2h_3 \w'/\w ) ] \,,
\nn
\vdots
\nn
g_{1+3}&=& \frac{\w^3 \csc 2\b }{2 f_4^{1/2} h_2 h_3}
[
2 (f_3f_4)^{1/2} h_1 [-h_2h_3+f_4 (k_2 h_3+k_3 h_2) ]+\cos 2\b [ (f_4h_1)' h_2 h_3\nn&&-f_4h_1 (h_2h_3)']] \,,
\nn
\vdots
\nn
g_7&=& \frac{\w^3}{h_1h_2h_3} \left[-(f_3f_4)^{1/2} [k_1h_2h_3+\mbox{perm}] +
\cos 2\b [(h_1h_2h_3)'+6 h_1h_2h_3 \w'/\w] \right] \,,\nn
\eea
\end{small}
with the appropriate permutations to obtain $g_2, g_3$ and $g_5, g_6$.
We then used these to express the quantities
\bea
\Big(\frac{g_4}{A_1^2}+\mbox{perm}\Big) \sac
\Big(\frac{g_3}{A_1^2A_2^2}+\mbox{perm}\Big),
\eea
in terms of the metric functions and their first derivatives. After
inserting these expressions into the remaining four Killing spinor
equations we obtain a system of four coupled first order ODEs for the
metric functions:
\begin{small}
\bea
0&=&
\frac{\omega^3}{12}\Big[\sin^22\beta\log(\w^{12}f_4^3h_1h_2h_3)'+\cos^22\b\log\Big(\frac{f_4^3}{h_1h_2h_3}\Big)'\Big]
\nn&& +\cot2\b\Big[\frac{f_4^{1/2}}{3}\sum_i\frac{k_i}{h_i}-\frac{1}{2f_4^{1/2}}\Big]
\,, \nn
1&=&\frac{\omega^3}{12}\Big[\sin^22\beta\log(\w^{12}f_4^3h_1h_2h_3)'+\cos^22\b\log\Big(\frac{f_4^3}{h_1h_2h_3}\Big)'\Big]
\nn&& +\cot2\b\Big[\frac{f_4^{1/2}}{3}\sum_i\frac{k_i}{h_i}-\frac{1}{2f_4^{1/2}\cos^22\b}\Big]
\,,\nn
0&=&(\w^3)'+\frac{\w^3}{2}(1+\cos^22\b)\log(\w^6h_1h_2h_3)'-f_4^{1/2}\cot2\b\sum_i\frac{k_i}{h_i}
\,, \nn
0&=&\w^3\Big[\sin2\b\b'+\frac{\cos2\b}{8}\log\Big(\frac{f^3_4}{h_1h_2h_3}\Big)'\Big]+\frac{1}{\sin2\b}\Big[\frac{f^{1/2}_4}{2}\sum_i\frac{k_i}{h_i}-\frac{3}{4f_4^{1/2}}\Big]-\frac{1}{2}
\,.\nonumber
\eea
\end{small}
These equations are not all
independent, and in fact reduce to one algebraic condition\footnote{Note that \bref{f4}, together with \bref{gauge},
imply that $\sqrt{f_3 f_4}=-1/2\w^3$. This sign is important, for example, when checking the equations of motion
for the four-form. \label{foot}}
\be
\label{f4}
f_4^{1/2}=-\frac{\sin2\b}{2} \,,
\ee
and two differential conditions:
\bea\label{twodiff}
\log(\w^{12}\sin^22\b h_1h_2h_3)'&=&-\frac{4\cos2\b}{\w^3\sin^22\b},\nn
(\w^3\cos2\b)'&=&2-g_7.
\eea
It is convenient, therefore, to define
\be
f = \w^3 \cos 2\b\,,
\ee
and trade $\b$ for $f$.

The next step is to integrate the Bianchi identities \bref{bianchis}. From the expressions for
the flux components one obtains
\bea
k_3g_5+k_2g_6&=&-\frac{1}{2}[f(k_2h_3+k_3h_2)+k_2k_3r]',\nn
g_1&=&-\frac{1}{2}f(k_2h_3+k_3h_2)-(\w^6h_2h_3)',
\eea
together with permutations of $(1,2,3)$. We may therefore
integrate the Bianchi identities twice to find
\bea
g_1 &=&-\undos \left[ f (k_2 h_3 + k_3 h_2) + k_2 k_3 r \right] + c_1\,,
\nn
g_2 &=&-\undos \left[ f (k_3 h_1 + k_1 h_3) + k_3 k_1 r \right] + c_2 \,,
\nn
g_3 &=&-\undos \left[ f (k_1 h_2 + k_2 h_1) + k_1 k_2 r \right] + c_3\,,
\label{bianchis2}
\eea
and
\bea
0&=& \w^6 h_1 h_2 - {1 \over 4} k_1 k_2 r^2 + c_3 r + d_3\,,
\nn
0&=& \w^6 h_2 h_3 - {1 \over 4} k_2 k_3 r^2 + c_1 r + d_1\,,
\nn
0&=& \w^6 h_3 h_1 - {1 \over 4} k_3 k_1 r^2 + c_2 r + d_2\,,
\label{amazing}
\eea
for some constants $c_i$, $d_i$. The expression for $g_7$ now takes the form
\be
g_7= {f_3 \over 2 h_1h_2h_3} \Big[ (\w^6+f^2) (k_1 h_2 h_3 +\mbox{perm.} )
+ f(k_1 k_2 h_3+\mbox{perm})r -2 fc_i h_i\Big]\,,
\ee
It thus remains to solve the two coupled first order equations \bref{twodiff}.
The function redefinitions
\be
H=\w^6 h_1h_2h_3 (\w^6-f^2)  \sac I=4 \w^6 h_1h_2h_3 f  \,,
\ee
allows us to decouple them, at the expense of having to solve one single second order differential equation
\begin{small}
\bea
\label{IH}
I &=&-H'
\,,\nn
0&=& -4 (H')^2+4 H(2 H''+4 k_i d_i + 4  k_i c_i r -3 k_1k_2k_3 r^2)
{} + \prod_{i=1}^3 \left({k_1k_2k_3 \over k_i} r^2 -4 r c_i -4d_i \right)
\,. \nn
\eea
\end{small}
From any solution of this second order ODE we can reconstruct the full solution. The explicit formulae
are recorded in the main text.


\subsection{G-structure analysis}
\label{sec:calib}

In this appendix we use the G-structure techniques of~\cite{GMMW1}
to derive, first, a set of general supersymmetry conditions for a
class of $AdS_3$ backgrounds, and then consider the restriction to the
specific ansatz~\eqref{an1}--\eqref{an3}. The conditions are derived
as a limit of class of warped $\bbR^{1,1}$ supersymmetric
Minkowski backgrounds. The discussion follows exactly that
in~\cite{GMMW1}, except that here we include an electric flux
component.

The class of Minkowski backgrounds is defined by the
set of projections on the Killing spinors. The solutions of interest
are related to M5-branes wrapping holomorphic four-cycles in a
Calabi--Yau fourfold (``4-in-8  K\"ahler'' solutions) with
$N=(0,2)$ supersymmetry.  We start by considering the geometry with the M5-brane
viewed as a probe in the special holonomy spacetime
$\bbR^{1,2}\times\mathcal{M}_{SU(4)}$. We can choose a frame
$\{e^+,e^-,e^1,\dots,e^9\}$, with $\bbR^{1,2}$ spanned by
$\{e^+,e^-,e^9\}$, such that the four special holonomy Killing spinors
satisfy
\begin{equation}
   \Gamma^{1234}\epsilon = \Gamma^{3456}\epsilon
      = \Gamma^{5678}\epsilon = - \epsilon \,.
\end{equation}
The $SU(4)$ structure can be written as
\begin{equation}
\begin{aligned}
   \tilde{J} &= e^{12} + e^{34} + e^{56} + e^{78} \,, \\
   \tilde{\Omega}
      &= (e^1+ie^2)\wedge(e^3+ie^4)\wedge(e^5+ie^6)\wedge(e^7+ie^8) \,,
\end{aligned}
\end{equation}
where $e^{i_1\dots i_n}=e^{i_1}\wedge\dots\wedge e^{i_n}$. The
projection on the probe brane can be written
$\Gamma^{+-1234}\epsilon=-\epsilon$ or equivalently
\begin{equation}
   \Gamma^{+-}\epsilon = \epsilon \,.
\end{equation}
Observe that these projections imply that $\G^{+-9}\e=\e$. Therefore,
we can add space-filling probe membranes without breaking any further
supersymmetry. These are sources for electric flux in the supergravity
description of the backreacted geometry. Together these projections define an $(SU(4)\ltimes\bbR^8)\times\bbR$
structure~\cite{oap2}, or equivalently, a pair of
$(Spin(7)\ltimes\bbR^8)\times\bbR$ structures
$K=e^+$, $\Omega_{M}=K\wedge v$, $\Sigma=K\wedge \phi$ with
$\phi=\phi_\pm$ and $v=e^9$ where
\begin{equation}
   \phi_\pm = - \tfrac{1}{2}\tilde{J}\wedge \tilde{J}
      \mp \re\tilde{\Omega} \,.
\end{equation}
We define a class of ``wrapped-brane'' geometries as warped
$\bbR^{1,1}\times\mathcal{M}_9$ backgrounds where the Killing spinors
satisfy the same projections as the probe brane geometry. The metric
is assumed to have the form
\begin{equation}
   ds^2 = L^{-1}ds^2(\bbR^{1,1}) + ds^2(\mathcal{M}_9) \,,
\end{equation}
so $e^+=L^{-1}du$ and $e^-=dv$, while, preserving the Minkowski
symmetries, we split the flux as
\begin{equation}
   G_4 = e^{+-}\wedge E_2 + B_4 \,.
\end{equation}
Note that unlike the discussion in~\cite{GMMW1} we keep some electric
flux $E_2$ as well as magnetic flux $B_4$.

A necessary condition for supersymmetry is that the
$(Spin(7)\ltimes\bbR^8)\times\bbR$ calibration
conditions~\cite{Gauntlett:2002fz} are satisfied, namely
\begin{equation}
\begin{aligned}
   dK &= \tfrac{2}{3}i_{\Omega_M} G_4
      + \tfrac{1}{3}i_\Sigma\star G_4 \,, \\
   d\Omega_M &= i_K G_4 \,, \\
   d\Sigma &= i_K\star G_4 - \Omega_M\wedge G_4 \,,
\end{aligned}
\end{equation}
Substituting for the
pair of structures $\phi_\pm$ one finds the conditions
\begin{equation}
\label{calibconds}
\begin{aligned}
   L^{-1}dL &=\tfrac{2}{3}i_vE_2
      - \tfrac{1}{6}i_{\tilde{J}\wedge\tilde{J}}\star_9B_4\,, \\
   \re\tilde{\Omega} \wedge B_4 &= 0 \,, \\
   Ld(L^{-1}v) &= E_2 \,, \\
   d(L^{-1}\re\tilde{\Omega}) &= 0 \,, \\
   \tfrac{1}{2}Ld(L^{-1}\tilde{J}\wedge \tilde{J})
      &= \star_9 B_4 - v\wedge B_4 \,,
\end{aligned}
\end{equation}
where the orientation on $\mathcal{M}_9$ is defined by
$\Vol{\mathcal{M}_9}=e^{1\dots9}$. Generically, since
the Killing spinors for supersymmetric spacetimes satisfying these
conditions define a preferred local $(SU(4)\ltimes\bbR^8)\times\bbR$
structure, given our metric ansatz one must impose
the Bianchi identity and the $+-9$ component of the four-form field
equations to ensure that all field equations are satisfied \cite{oap3}.

To obtain conditions for supersymmetric $AdS_3\times\mathcal{M}_8$
spacetimes, one specialises the conditions~\eqref{calibconds} by
assuming the warping and metric have the form
\begin{equation}
\begin{aligned}
   L &= e^{2R}\omega^{-2} \,, \\
   ds^2(\mathcal{M}_9) &=
      \omega^2 \left[ dR^2 + ds^2(\mathcal{M}_8) \right] \,.
\end{aligned}
\end{equation}
This matches the ansatz~\eqref{an1} with the radial coordinate $R$
combining with the $\bbR^{1,1}$ factor to give a unit radius $AdS_3$
in Poincar\'e coordinates. To preserve the $AdS_3$ symmetries one
assumes in addition that
\begin{equation}
   E_2 = \omega dR \wedge E_1
\end{equation}
and that $B_4$ has no component along $dR$.

In general the radial direction will only lie partly along
$v$. If the orthogonal component lies along a unit one form $\hat{u}$
then writing $v=\omega\hat{v}$ we have
\begin{equation}
\begin{aligned}
   dR &= \cos2\beta\, \hat{v} + \sin2\beta\, \hat{u} \,, \\
   \hat{\rho} &= - \sin2\beta\, \hat{v} + \cos2\beta\, \hat{u} \,,
\end{aligned}
\end{equation}
where we have also introduced $\hat{\rho}$, the unit one-form
orthogonal to $dR$. Note that the angle $\beta$ is the same angle
defined in the Killing spinor analysis above. This decomposition
defines a (local) $SU(3)$ structure on $\mathcal{M}_8$, defined by
$\hat{\rho}$, together with a second unit one-form $\hat{w}=J\hat{u}$,
a two-form $J$ and three-form $\Omega$. The relation to the original
$SU(4)$ structure is
\begin{equation}
\label{reduct}
\begin{aligned}
   \omega^{-2}\tilde{J} &= J + \hat{w} \wedge \hat{u} \,, \\
   \omega^{-4}\tilde{\Omega} &= \Omega \wedge (\hat{w} + i\hat{u}) \,.
\end{aligned}
\end{equation}
The metric on $\mathcal{M}_8$ can be written as
\begin{equation}
   ds^2(\mathcal{M}_8) = ds^2(\mathcal{M}_{SU(3)})
      + \hat{\rho}^2 + \hat{w}^2 \,,
\end{equation}
where $ds^2(\mathcal{M}_{SU(3)})$ is the metric defined by the $SU(3)$
structure $(J,\Omega)$.

Given the relations~\eqref{reduct}, reducing the
conditions~\eqref{calibconds} one finds
\begin{align}
   \label{gaugeeq}
   d(\omega^3\sin2\beta\,\hat{\rho}) &= 0 \,, \\
   \label{Omeq}
   d(\omega^6\sin2\beta\,\im\Omega) &
      = -2\omega^6\left(\re\Omega\wedge\hat{w} -
      \cos2\beta\,\im\Omega\wedge\hat{\rho}\right)
\end{align}
together with
\begin{equation}
\label{calflux}
\begin{gathered}
   d(\omega^3\cos2\beta) - 2\omega^3\sin2\beta\,\hat{\rho}
       = - \omega^3 E_1 \,, \\
   6\omega^{-1} d\omega
       = -2\cos2\b E_1 + \omega^{-3}\sin2\b i_{J\wedge\hat{w}}\star_8B_4 \,, \\
   d\left[\omega^6\left(\tfrac12 J\wedge J
          + \cos2\beta\, J\wedge \hat{w}\wedge\hat{\rho}\right)\right]
       = \omega^3 \sin2\beta\, \hat{\rho}\wedge B_4 \,, \\
   \begin{aligned}
   - d\left(\omega^6\sin2\beta\, J\wedge\hat{w}\right)
         +& 2\omega^6\left(\tfrac12 J\wedge J
         + \cos2\beta\, J\wedge\hat{w}\wedge\hat{\rho}\right) \\
      &\qquad\qquad\qquad\qquad
      = \omega^3 \star_8 B_4 + \omega^3 \cos2\beta\, B_4 \,,
   \end{aligned}
\end{gathered}
\end{equation}
and the algebraic constraints
\begin{equation}
\begin{aligned}
   \im\Omega\wedge B_4 &= 0 \,, \\
   \hat{w} \wedge \re\Omega \wedge B_4 &= 0 \,, \\
   4\omega^3\sin2\b i_{\hat{\rho}}E_1
      + \left(i_{J\wedge J}
         + 2\cos2\b\,i_{J\wedge\hat{w}\wedge\hat{\rho}}\right)\star_8B_4
      &= 12\omega^3 \,,
\end{aligned}
\end{equation}
where the orientation on $\mathcal{M}_8$ is given by
$\Vol{\mathcal{M}_8}=\tfrac{1}{6}J\wedge J\wedge J\wedge
\hat{\rho}\wedge\hat{w}$. Note that this is not necessarily a minimal
set of conditions: one expects that there is redundancy among these
relations.

In addition, one must impose the Bianchi
identity for $G_4$, which, given~\eqref{calflux} requires that we impose
\begin{equation}
\label{bi}
   dB_4 = 0 \,.
\end{equation}
For the specialisation to $AdS_3$, the preferred
local structure group defined by the Killing spinors of the wrapped
brane spacetime is reduced from $(SU(4)\ltimes\bbR^8)\times\bbR$ to
$SU(3)$. We may read off from the results of~\cite{oap} that for the
$AdS_3$ spacetimes it is sufficient to impose just the Bianchi
identity in addition to the supersymmetry conditions, and all field
equations are then identically satisfied.

Let us now compare our general conditions with the specific
ansatz~\eqref{an1}--\eqref{an3} relevant to our solutions. The
identification is
\begin{equation}
\begin{aligned}
   J &= h_1 J_1 + h_2 J_2 + h_3 J_3 \,, \\
   \Omega
      &= (h_1h_2h_3)^{1/2} \Omega_1\wedge\Omega_2\wedge\Omega_3 \,, \\
   \hat{\rho} &= f_3^{1/2} dy \,, \\
   \hat{w} &= f_4^{1/2} D\psi \,,
\end{aligned}
\end{equation}
where $\Omega_i$ are the holomorphic one-forms on K\"ahler-Einstein
spaces $C_i$. Substituting into the
conditions~\eqref{gaugeeq}--\eqref{bi} one can relatively quickly
derive the supersymmetry conditions given in the previous
section. Rather than repeat that calculation in detail, let us make a
couple of observations. From~\eqref{gaugeeq}, one notes that, quite
generically for any $AdS_3$ solution, one can introduce a coordinate $y$
such that
\begin{equation}
   \hat{\rho} = \frac{dy}{\omega^3\sin2\beta} \,.
\end{equation}
This is precisely the gauge condition~\eqref{gauge} we chose in
analysing the Killing spinor equation and was the motivation for this
choice. Using this gauge, it is easy to see, for instance, that the
$\hat{\rho}$ component of~\eqref{Omeq} gives the first differential
equation in~\eqref{IH}.

\section{Some properties of $KE_4$ spaces}
\label{sec:appendix-KE4}

In the main text we make use of a few properties of four-dimensional
K\"ahler--Einstein spaces. Let us summarise and explain them
here. For some more details see
refs.~\cite{GH,Beauville,DLOW}

Given a K\"ahler metric on a complex manifold $\mathcal{M}$ one can
always construct the Ricci form $\mathcal{R}$ by contracting the
Riemann tensor with $J$. This two-form gives the curvature of a
holomorphic line bundle $\mathcal{L}$ known as the anti-canonical
bundle. The first Chern class $c_1(\mathcal{M})$ of $\mathcal{M}$ is
just the first Chern class of $\mathcal{L}$. It is given by
$\frac{1}{2\pi}\mathcal{R}$, is necessarily integral and depends only on
the choice of complex structure on $\mathcal{M}$. In the special case
where the metric is also Einstein it is easy to show that the Ricci
form is proportional to the K\"ahler form $J$,
\begin{equation}
   \mathcal{R} = k J \,.
\end{equation}
In this paper we normalise the metric such that $k=0,\pm1$. If $k=0$ the
metric is Calabi--Yau.

Suppose we have a minimal set of two-cycles $\{\Sigma_a\}$ which
generate $H_2(\mathcal{M},\bbZ)$. This means that any element of the
homology can be written as $m^a\Sigma_a$ for some set of integers
$m^a$. It is possible that $H_2(\mathcal{M},\bbZ)$ includes torsion
elements. These are elements $\Sigma$ of such that $\Sigma$ itself is
non-trivial but $p\Sigma=0$ in cohomology for some $p\in\bbZ$. If we
ignore this subtlety (in fact none of the manifolds we are interested
in will have torsion) then the integrality of $c_1(\mathcal{M})$ is
equivalent to the integrality of the periods
\begin{equation}
   n(\Sigma_a) = \int_{\Sigma_a} c_1(\mathcal{M})
      = \frac{1}{2\pi}\int_{\Sigma_a}\mathcal{R} \in \bbZ \,.
\end{equation}
In some cases, the anti-canonical bundle can be written as a
(positive) multiple tensor product of another line bundle, that is
$\mathcal{L}=\mathcal{N}^m$ for $m\in\bbZ^+$. In general the maximal possible
$m$ is known as the Fano index. It implies that
$c_1(\mathcal{M})=mc_1(\mathcal{N})$ and hence $n(\Sigma_a)=mn_a$
where $n_a=\int_{\Sigma_a}c_1(\mathcal{N})$. Since $m$ is maximal, the
$n_a$ must be coprime.

Recall that, in $d$ dimensions, $c_1(\mathcal{M})\in
H^2(\mathcal{M},\bbZ)$ is Poincar\'e dual to a $(d-2)$-cycle
$\Sigma_{\mathcal{L}}$ in $H_{d-2}(\mathcal{M},\bbZ)$. This means by
definition that the integers $n(\Sigma_a)$ are given by the
intersection number of $\Sigma_{\mathcal{L}}$ with $\Sigma_a$
\begin{equation}
   n(\Sigma_a) = \Sigma_{\mathcal{L}} \cdot \Sigma_a \,.
\end{equation}
Note that in $d=4$, $\Sigma_{\mathcal{L}}\in H_2(\mathcal{M},\bbZ)$ and so
can be written as $\Sigma_{\mathcal{L}}=ms^a\Sigma_a$ for some set of
integers $s^a\in\bbZ$.

Of specific interest here are those four-dimensional complex
manifolds $KE_4^+$ which admit a positive-curvature K\"ahler--Einstein
metric. The list is in fact finite: only $CP^2$, $S^2\times S^2$ and
the del Pezzo surfaces $\dP_k$ for $k=3,\dots,8$ are allowed. These
latter spaces are $CP^2$ blown up at $k$ distinct points. None of
these spaces have torsion classes. In four dimensions, in addition to
the Fano index $m$ and the periods $n_a$, one can also
consider the integer
\begin{equation}
   M = \int_{KE^+_4} c_1(KE^+_4) \wedge c_1(KE^+_4)
      = m^2 \int_{KE^+_4} c_1(\mathcal{N}) \wedge c_1(\mathcal{N})
      = \Sigma_{\mathcal{L}} \cdot \Sigma_{\mathcal{L}}
      \in \bbZ \,.
\end{equation}
Note that by construction $M$ is always divisible by $m^2$. These are
all the quantities which will enter our discussion.

For completeness, it is straightforward to list $m$, $\{n_a\}$ and $M$
for each of the allowed $KE^+_4$ spaces. For $CP^2$ the situation is
very simple: $H_2(CP^2,\bbZ)=\bbZ$ and is generated by a single class,
known as the hyperplane class $H$. A representative curve in this
class is just the $S^2$ given by say $z_1=0$ in homogeneous
coordinates $(z_1,z_2,z_3)$. It is easy to see that $H\cdot H=1$. One
finds that $\Sigma_{\mathcal{L}}=3H$ and so $m=3$, $n_1=1$ and $M=9$.

If $KE^+_4=S^2\times S^2$ things are equally straightforward. Now
$H_2(S^2\times S^2,\bbZ)=\bbZ^2$, and is simply generated by the
classes $H_1$ and $H_2$ corresponding to each of the two
spheres. The only non-zero intersection is $H_1\cdot H_2=1$. One finds
$\Sigma_{\mathcal{L}}=2(H_1+H_2)$ and so $m=2$, $n_1=n_2=1$ and $M=8$.

For $\dP_k$ recall that in four dimensions blowing up replaces
a point on the manifold with a two-sphere. Thus
$H_2(\dP_k,\bbZ)=\bbZ^{k+1}$ and is generated by $H$, the image of the
hyperplane class after blowing up $CP^2$, together with the
exceptional two-spheres $E_i$, for $i=1,\dots,k$, at the $k$ blown-up
points. The only non-zero intersections are $H\cdot H=1$ and $E_i\cdot
E_j=-\delta_{ij}$. One finds $\Sigma_{\mathcal{L}}=3H-E_1-\dots-E_k$ and so
$m=1$ while $n_i=1$ and $n_{k+1}=3$ (if we label $\Sigma_i=E_i$ and
$\Sigma_{k+1}=H$). In addition $M=9-k$.

\section{Reduction to type IIB}
\label{sec:appendix-orientations}

Consider a solution of eleven-dimensional supergravity invariant under
a $T^2$ action of the form
\bea
ds^2_{11}&=&ds^2_9+f^2 ds^2(T^2)
\,,\nn
G_4 &=& F_4 + F_2 \wedge \Vol{T^2}\,,
\eea
with orientation
\bea
\e_{11}&=&\e_9 \wedge \Vol{T^2} \,,
\eea
where $\e_9$ defines an orientation on $ds^2_9$. The Bianchi identity,
$dG_4=0$, and the eleven-dimensional equation of motion for the
four-form, $d\star G_4+{1
  \over 2} G_4 \wedge G_4 =0$, decompose as
\begin{equation}
\label{po}
\begin{gathered}
   dF_2 = 0 \,, \qquad \qquad \qquad  dF_4 = 0 \, \\
   d\left( f^2 \star_9 F_4 \right) +F_2 \wedge F_4 = 0 \,, \qquad
   d\left( f^{-2} \star_9 F_2 \right) +\tfrac{1}{2}F_4 \wedge F_4
      = 0 \,,
\end{gathered}
\end{equation}
where $\star_9$ is the Hodge dual with respect to $ds^2_9$.

After dimensional reduction and T-duality, we find the following
metric and five-form for type IIB supergravity
\bea
ds^2_{10} &=& f ds^2_9 + {1 \over f^3} (dz+ A_1)^2 
\,, \nn
F_5&=&(1+\star) F_4\wedge (dz+ A_1) \,,\nn
   &=&F_4\wedge (dz+ A_1)+f^2\star_9 F_4
\eea
with $dA_1=F_2$. Observe that the orientation in ten dimensions is
given by
\be
\e_{10} = \e_9 \wedge dz \,,
\ee
so that closure of $F_5$ indeed follows from (\ref{po}).

It is also useful to note here the standard relations between
eleven-dimensional M-theory parameters and the ten-dimensional type
IIB parameters~\cite{M-IIB}. Following
Polchinski~\cite{Pol}, we define the eleven-dimensional Planck length
$l_P$ by  $2\kappa^2_{11}=(2\pi)^8l_P^9$. For a square torus, with sides
length $2\pi R_1$ and $2\pi R_2$ the string length $l_s$, and type IIB
coupling are given by
\begin{equation}
   l_s^2 = l_P^3/R_1 \,, \qquad
   g_s = R_1/R_2 \,.
\end{equation}
The radius $R_\mathrm{IIB}$ of the IIB circle is given by
\begin{equation}
   R_\mathrm{IIB} = l_P^3/R_1R_2 \,.
\end{equation}




\begin{thebibliography}{99}

\bibitem{Maldacena:1997re}
  J.~M.~Maldacena,
  ``The large N limit of superconformal field theories and supergravity,''
  Adv.\ Theor.\ Math.\ Phys.\  {\bf 2} (1998) 231
  [Int.\ J.\ Theor.\ Phys.\  {\bf 38} (1999) 1113]
  [arXiv:hep-th/9711200].

\bibitem{GMMW2}
  J.~P.~Gauntlett, O.~A.~P.~Conamhna, T.~Mateos and D.~Waldram,
   ``Supersymmetric $AdS_3$ solutions of type IIB supergravity'', to
   appear in Phys.Rev.Lett, arXiv:hep-th/0606221.

\bibitem{GMSW}
  J.~P.~Gauntlett, D.~Martelli, J.~Sparks and D.~Waldram,
  ``Supersymmetric $AdS_5$ solutions of M-theory,''
  Class.\ Quant.\ Grav.\  {\bf 21} (2004) 4335
  [arXiv:hep-th/0402153].

\bibitem{GMSW2}
  J.~P.~Gauntlett, D.~Martelli, J.~Sparks and D.~Waldram,
  ``Sasaki-Einstein metrics on $S_2\times S_3$,''
  Adv.\ Theor.\ Math.\ Phys.\  {\bf 8} (2004) 711
  [arXiv:hep-th/0403002].

\bibitem{Cvetic:2005ft}
  M.~Cvetic, H.~Lu, D.~N.~Page and C.~N.~Pope,
  ``New Einstein-Sasaki spaces in five and higher dimensions,''
  Phys.\ Rev.\ Lett.\  {\bf 95} (2005) 071101
  [arXiv:hep-th/0504225].

\bibitem{gauge}
  M.~Bertolini, F.~Bigazzi and A.~L.~Cotrone,
   ``New checks and subtleties for AdS-CFT and a-maximization,''
  JHEP {\bf 0412} (2004) 024
  [arXiv:hep-th/0411249].

\bibitem{gauge2}
  S.~Benvenuti, S.~Franco, A.~Hanany, D.~Martelli and J.~Sparks,
  ``An infinite family of superconformal quiver gauge theories with
  Sasaki-Einstein duals,''
  JHEP {\bf 0506} (2005) 064
  [arXiv:hep-th/0411264].

\bibitem{gauge3}
  S.~Benvenuti and M.~Kruczenski,
   ``From Sasaki-Einstein spaces to quivers via BPS geodesics: $L(p,q|r)$,''
  %
  JHEP {\bf 0604} (2006) 033
  [arXiv:hep-th/0505206].

\bibitem{gauge4}
  S.~Franco, A.~Hanany, D.~Martelli, J.~Sparks, D.~Vegh and B.~Wecht,
   ``Gauge theories from toric geometry and brane tilings,''
  %
  JHEP {\bf 0601} (2006) 128
  [arXiv:hep-th/0505211].

\bibitem{gauge5}
  A.~Butti, D.~Forcella and A.~Zaffaroni,
   ``The dual superconformal theory for $L(p,q,r)$ manifolds,''
  JHEP {\bf 0509} (2005) 018
  [arXiv:hep-th/0505220].


\bibitem{Gauntlett:2005ww}
  J.~P.~Gauntlett, D.~Martelli, J.~Sparks and D.~Waldram,
  ``Supersymmetric $AdS_5$ solutions of type IIB supergravity,''
  arXiv:hep-th/0510125.

\bibitem{ruben}
  R.~Minasian, M.~Petrini and A.~Zaffaroni,
   ``Gravity duals to deformed SYM theories and generalized complex geometry,''
  %
  arXiv:hep-th/0606257.

\bibitem{Pilch:2000ej}
  K.~Pilch and N.~P.~Warner,
  ``A new supersymmetric compactification of chiral IIB supergravity,''
  Phys.\ Lett.\ B {\bf 487} (2000) 22
  [arXiv:hep-th/0002192].

\bibitem{GMMW1}
  J.~P.~Gauntlett, O.~A.~P.~Mac Conamhna, T.~Mateos and D.~Waldram,
  ``AdS spacetimes from wrapped M5 branes'', to appear in JHEP,
  arXiv:hep-th/0605146. 


\bibitem{naka}
  M.~Naka,
  ``Various wrapped branes from gauged supergravities,''
  arXiv:hep-th/0206141.

\bibitem{mn}
  J.~M.~Maldacena and C.~Nunez,
  ``Supergravity description of field theories on curved manifolds and a no go
  theorem,''
  Int.\ J.\ Mod.\ Phys.\ A {\bf 16} (2001) 822
  [arXiv:hep-th/0007018].

\bibitem{GKW}
  J.~P.~Gauntlett, N.~Kim and D.~Waldram,
  ``M-fivebranes wrapped on supersymmetric cycles,''
  Phys.\ Rev.\ D {\bf 63} (2001) 126001
  [arXiv:hep-th/0012195].

\bibitem{wit}
  E.~Witten,
 ``On flux quantization in M-theory and the effective action,''
  J.\ Geom.\ Phys.\  {\bf 22} (1997) 1
  [arXiv:hep-th/9609122].

\bibitem{GH}
  P.~Griffiths and J.~Harris,
  Principles of Algebraic Geometry,
  John Wiley and Sons, New York, 1994.

\bibitem{Beauville}
  A.~Beauville,
  Complex Algebraic Surfaces,
  LMS Student Texts 34, CUP, Cambridge, UK, 1996

\bibitem{Gauntlett:2001qs}
  J.~P.~Gauntlett, N.~Kim, S.~Pakis and D.~Waldram,
  ``Membranes wrapped on holomorphic curves,''
  Phys.\ Rev.\ D {\bf 65} (2002) 026003
  [arXiv:hep-th/0105250].

\bibitem{Brown:1986nw}
  J.~D.~Brown and M.~Henneaux,
  ``Central Charges In The Canonical Realization Of Asymptotic Symmetries: An
  Example From Three-Dimensional Gravity,''
  Commun.\ Math.\ Phys.\  {\bf 104} (1986) 207.

\bibitem{kr}
  A.~Karch and L.~Randall,
  ``Locally localized gravity,''
  JHEP {\bf 0105} (2001) 008
  [arXiv:hep-th/0011156].

\bibitem{st}
  K.~Skenderis and M.~Taylor,
  ``Branes in AdS and pp-wave spacetimes,''
  JHEP {\bf 0206} (2002) 025
  [arXiv:hep-th/0204054].

\bibitem{Gauntlett:2002fz}
  J.~P.~Gauntlett and S.~Pakis,
  ``The geometry of $D=11$ Killing spinors,''
  JHEP {\bf 0304} (2003) 039
  [arXiv:hep-th/0212008].

\bibitem{oap} O. A. P. Mac Conamhna, ``The geometry of extended null
  supersymmetry in M-theory'', Phys.Rev. D{\bf 73} (2006) 045012,
  hep-th/0505230.

\bibitem{oap2} M. Cariglia and O. A. P. Mac Conamhna, ``Null structure
  groups in eleven dimensions'', Phys.Rev. D{\bf 73} (2006) 045011,
  hep-th/0411079.

\bibitem{oap3} O. A. P. Mac Conamhna, ``Eight-manifolds with
  G-structure in eleven dimensional supergravity'', Phys.Rev. D{\bf 72}
  (2005) 086007, hep-th/0504028.

\bibitem{Lu:1998nu}
  H.~Lu, C.~N.~Pope and J.~Rahmfeld,
  ``A construction of Killing spinors on $S^n$,''
  J.\ Math.\ Phys.\  {\bf 40} (1999) 4518
  [arXiv:hep-th/9805151].

\bibitem{DLOW}
  R.~Donagi, A.~Lukas, B.~A.~Ovrut and D.~Waldram,
  ``Holomorphic vector bundles and non-perturbative vacua in M-theory,''
  JHEP {\bf 9906}, 034 (1999)
  [arXiv:hep-th/9901009].

\bibitem{M-IIB}
   J.~H.~Schwarz,
   ``An $SL(2,\bbZ)$ multiplet of type IIB superstrings,''
   Phys.\ Lett.\ B {\bf 360}, 13 (1995)
   [Erratum-ibid.\ B {\bf 364}, 252 (1995)]
   [arXiv:hep-th/9508143], \\
   P.~S.~Aspinwall,
   ``Some relationships between dualities in string theory,''
   Nucl.\ Phys.\ Proc.\ Suppl.\  {\bf 46}, 30 (1996)
   [arXiv:hep-th/9508154].

\bibitem{Pol}
   J~Polchinski,
   ``String Theory, vol. 2: Superstring theory and beyond'',
   Cambridge University Press, UK (1998).


\end{thebibliography}
\end{document}